\documentclass[fleqn,usenatbib]{mnras}

\usepackage[T1]{fontenc}

\DeclareRobustCommand{\VAN}[3]{#2}
\let\VANthebibliography\thebibliography
\def\thebibliography{\DeclareRobustCommand{\VAN}[3]{##3}\VANthebibliography}

\usepackage{graphicx}	
\usepackage{amsmath}	
\usepackage{amssymb}	

\usepackage{newtxtext,newtxmath}

\newcommand{\mdot}{$\mathrm{M}_\odot$}
\newcommand{\zdot}{$\mathrm{Z}_\odot$}
\newcommand{\healpix}{\textsc{healpix}}
\newcommand{\treecol}{\textsc{treecol}}

\title[]{Formation of star clusters and enrichment by massive stars in simulations of low-metallicity galaxies with a fully sampled initial stellar mass function}

\author[Lah\'en et al.]{
Natalia Lah\'en$^{1}$\thanks{E-mail: nlahen@mpa-garching.mpg.de},
Thorsten Naab$^{1}$, 
Guinevere Kauffmann$^{1}$, 
Dorottya Sz\'ecsi$^{2}$,
Jessica May Hislop$^{3,1}$, \newauthor
Antti Rantala$^{1}$,
Alexandra Kozyreva$^{4}$,
Stefanie Walch$^{5}$ and 
Chia-Yu Hu$^{6,7}$
\vspace{1mm}
\\
$^{1}$Max-Planck-Institute f\"{u}r Astrophysik, Karl-Schwarzschild-Stra$\beta$e 1, D-85740 Garching, Germany\\
$^{2}$Institute of Astronomy, Faculty of Physics, Astronomy and Informatics, Nicolaus Copernicus University, Grudzi\k{a}dzka 5, 87-100 Toru\'n, Poland\\
$^{3}$Department of Physics, University of Helsinki, Gustaf Hällströmin katu 2, FI-00014, Helsinki, Finland\\
$^{4}$Heidelberger Institut f\"{u}r Theoretische Studien, Schloss-Wolfsbrunnenweg 35, 69118 Heidelberg, Germany\\
$^{5}$I. Physikalisches Institut, Universitat zu K\"{o}ln, Z\"{u}lpicher Str. 77, D-50937 K\"{o}ln, Germany\\
$^{6}$Max-Planck-Institut f\"{u}r Extraterrestrische Physik, Giessenbachstrasse 1, D-85748 Garching, Germany \\
$^{7}$Department of Astronomy, University of Florida, 211 Bryant Space Science Center, Gainesville, FL 32611, USA
}

\date{Accepted 12 April 2023. Received 10 March 2023; in original form 28 November 2022}

\pubyear{2021}

\begin{document}
\label{firstpage}
\pagerange{\pageref{firstpage}--\pageref{lastpage}}
\maketitle

\begin{abstract}

We present new \textsc{griffin} project hydrodynamical simulations that model the formation of galactic star cluster populations in low-metallicity (\mbox{$Z=0.00021$}) dwarf galaxies, including radiation, supernova and stellar wind feedback of individual massive stars. In the simulations, stars are sampled from the stellar initial mass function (IMF) down to the hydrogen burning limit of 0.08 M$_\odot$. Mass conservation is enforced within a radius of 1 pc for the formation of massive stars. We find that massive stars are preferentially found in star clusters and follow a correlation set at birth between the highest initial stellar mass and the star cluster mass that differs from pure stochastic IMF sampling. With a fully sampled IMF, star clusters lose mass in the galactic tidal field according to mass-loss rates observed in nearby galaxies. Of the released stellar feedback, 60\% of the supernova material and up to 35\% of the wind material reside either in the hot interstellar medium (ISM) or in gaseous, metal enriched outflows. While stellar winds (instantaneously) and supernovae (delayed) start enriching the ISM right after the first massive stars form, the formation of supernova-enriched stars and star clusters is significantly delayed (by $>50$ Myr) compared to the formation of stars and star clusters enriched by stellar winds. Overall, supernova ejecta dominate the enrichment by mass, while the number of enriched stars is determined by continuous stellar winds. These results present a concept for the formation of chemically distinct populations of stars in bound star clusters, reminiscent of multiple populations in globular clusters.

\end{abstract}




\section{Introduction}

Increasing evidence from observations of young massive star clusters in the Local Group indicate that they harbour very massive stars of more than $100$ \mdot{} (\citealt{2008A&A...478..219M, 2008MNRAS.389L..38S, 2016MNRAS.458..624C, 2018Sci...359...69S}), that are characterized by extreme luminosities,  wind mass-loss rates and wind velocities (\citealt{2007A&A...465.1003M, 2022A&A...663A..36B}).
The highest mass stars are found in the Large Magellanic Cloud, in line with arguments that lower metallicity environments would be more favourable for massive star formation (e.g. \citealt{2012MNRAS.422.2246M}). Young star clusters have also been observed to follow a correlation between the mass of the most massive star and the total cluster mass \citep{2003ASPC..287...65L, 2006MNRAS.365.1333W}. Due to the large amount of energy and momentum released via various channels of feedback throughout their lifetime, massive stars help to regulate the non-equilibrium evolution of the interstellar medium (ISM, \citealt{2004RvMP...76..125M, 2007ARA&A..45..565M}). Winds and radiation output of massive stars, especially at photoionizing energies \citep{2009ApJ...694L..26G, 2012MNRAS.424..377D, 2012MNRAS.427..625W, 2017MNRAS.466.3293P, 2018ApJ...859...68K}, set the stage for supernovae (SNe) that in turn drive galactic outflows often observed in galaxies (e.g. \citealt{1990ApJS...74..833H, 1999ApJ...513..156M}). Stellar winds provide continuous energy input into the surrounding ISM but produce only a small fraction of the integrated energy output of one single massive star \citep{2001PASP..113..677C, 2016MNRAS.458.3528H}. Stellar winds are therefore generally thought to play a minor role in the self-regulating life cycle of the ISM as the kinetic energy imparted by the winds is mostly negligible \citep{2021MNRAS.501.1352G} and confined within the HII region \citep{2001PASP..113..677C}. Additionally, winds may help carve lower density channels that act as escape routes for matter and radiation \citep{2009ApJ...703.1352K, 2013MNRAS.431.1337R, 2017MNRAS.464.3536R}.

However, in some environments such as the earliest embedded phase of star formation, in the densest regions of the ISM, or around young massive star clusters, the effect of photoionizing radiation can be diminished in comparison to stellar winds \citep{2002ApJ...566..302M, 2014MNRAS.442..694D, 2018MNRAS.478.4799H}. While the stellar wind velocities can reach thousands of km s$^{-1}$ \citep{2001A&A...369..574V}, the highest mass-loss rates of $<10^{-4}$ \mdot{} yr${^{-1}}$ enable the most massive stars to inject more than the equivalent of one core-collapse SN in integrated energy across their lifetimes \citep{2017MNRAS.466.1903G}. One of the major roles of stellar winds of massive stars is to provide a form of continuous chemical enrichment that acts before the disruptive SN explosions. Stellar winds by massive stars offer a compelling scenario for self-enrichment, especially in dense star cluster-forming regions where they may play a more central role \citep{2019ApJ...871...20S}. Numerical work on the ISM enrichment via stellar winds on the scales of molecular clouds and single isolated clusters have indicated that it may be possible to retain some of the stellar wind material even in the extreme radiation environment of young massive star clusters \citep{2019ApJ...871...20S, 2021ApJ...922L...3L}. These small scale simulations concentrated on the wind enrichment but excluded the SN feedback that is important on galactic scales. Other recent numerical studies, such as \citet{2013ApJ...770...25A}, \citet{2018MNRAS.480..800H}, \citet{2019MNRAS.482.1304E}, \citet{2022MNRAS.510.5592A}, \citet{2022MNRAS.516.5914C} and \citet{2023MNRAS.521.2196A}, have also included stellar wind feedback at varying detail but did not either specifically concentrate on star clusters or have not studied the detailed enrichment by SNe and winds.

We present new hydrodynamical simulations of dwarf galaxies that introduce an implementation for stellar tracks of very massive stars into the \textsc{sphgal} code \citep{2014MNRAS.443.1173H, 2016MNRAS.458.3528H, 2017MNRAS.471.2151H}. The simulations are a part of the Galaxy Realizations Including Feedback From INdividual massive stars (\textsc{griffin})\footnote{\url{https://wwwmpa.mpa-garching.mpg.de/~naab/griffin-project}} project which utilizes high-resolution hydrodynamical simulations to address current challenges in galaxy formation and evolution \citep{2017ARA&A..55...59N}. In the project we have previously studied the formation of star clusters and globular clusters \citep{2019ApJ...879L..18L, 2020ApJ...891....2L, 2020ApJ...904...71L, 2022MNRAS.509.5938H}, the impact of runaway stars on the galactic properties \citep{2022arXiv220509774S}, the $H_2$ distribution \citep{2022MNRAS.512.4736S} and the origin of the [CII]-emission \citep{2022ApJ...934..115B} in low-metallicity dwarf galaxies. The simulations have so far implemented feedback from single stars by including radiation and feedback by SNe and asymptotic giant branch (AGB) stars. As we showed in \citet{2022MNRAS.509.5938H}, the star formation implementation can heavily affect the properties of the star clusters forming in the simulations. Here we implement a star formation prescription that does not include any efficiency parameters \citep{1959ApJ...129..243S} and instead combines a Jeans-mass dependent threshold and a physically motivated time delay for early stellar feedback. Additionally, we allow the sampling of a full IMF between the hydrogen burning limit of 0.08 \mdot{} and 500 \mdot, according to the upper limit of stellar masses in the stellar tracks, taking local conservation of the available gas mass reservoir into account. 

The new star formation and wind models are introduced in Section \ref{section:simulations}, the properties of the star clusters formed in the dwarf galaxy simulation are presented in Section \ref{section:star_clusters}, and the star formation and IMF properties are analysed in Section \ref{section:star_formation}. 
The wind and SN enrichment, as well as the fate of material released via both channels, are shown in Section \ref{section:enrichment}. A summary of the results and an outlook are given in Section \ref{section:summary}.

\section{Simulations}\label{section:simulations}

We use the \textsc{sphgal} code, a modified version of the tree-smoothed particle hydrodynamics (SPH) code \textsc{gadget-3} \citep{2005MNRAS.364.1105S}, to perform the simulations. The original implementation of \textsc{sphgal} with significant improvements to the performance of SPH has been introduced in \citet{2014MNRAS.443.1173H}, \citet{2016MNRAS.458.3528H} and \citet{2017MNRAS.471.2151H}. The code has since been used in various high-resolution simulations of isolated and merging dwarf galaxies (e.g. \citealt{2020ApJ...891....2L, 2022MNRAS.509.5938H}). Here we introduce a modified version of the star formation and IMF-sampling routines presented in \citet{2016MNRAS.458.3528H}, \citet{2019MNRAS.483.3363H} and \citet{2020ApJ...891....2L}, and our new implementation of stellar winds and stellar tracks that provide the stellar wind and radiation rates released by individual massive stars.

\subsection{Cooling and star formation}\label{section:sf}
\textsc{sphgal} models the non-equilibrium cooling of gas between temperatures of 10 K and $3\times10^4$ K using a chemical network with six chemical species (H$_2$, H$^+$, H, CO, C$^+$, O) and free electrons as outlined in detail in \citet{2016MNRAS.458.3528H} and \citet{2017MNRAS.471.2151H}. High-temperature gas cools according to the metallicity dependent cooling tables of \citet{2009MNRAS.393...99W}.

Star formation proceeds in regions where the local Jeans-mass is resolved with less than half of the SPH kernel mass (\mbox{$M_J<200 $ \mdot}) at an effective star formation efficiency per free-fall time of 100\%. We follow the locally mass-conserving IMF sampling routine described in \citet{2019MNRAS.483.3363H} with a modified feedback time-delay and a fully sampled stellar IMF. All gas particles crossing the threshold are decoupled from the hydrodynamics and tagged as reservoir particles which will be utilized in the sampling of the IMF. To model the unresolved final collapse of gas into feedback-releasing stars, we implement a physically motivated time delay for the onset of stellar feedback. In analytic models, observations and high-resolution ISM simulations, the star formation time scale is often connected with the local dynamical time ($t_\mathrm{dyn}$) or the free fall time (e.g. \citealt{2005ApJ...630..250K, 2011ApJ...730...40P, 2012ApJ...745...69K, 2012ApJ...761..156F, 2020ApJ...898...52M}). We therefore store the local dynamical time scale of the star-forming gas given by $t_\mathrm{dyn}=(4\pi G \rho)^{-1/2}$, where $G$ is the gravitational constant and $\rho$ is the gas density, for each new reservoir particle at the moment the particle crosses the star formation threshold. These particles remain inert for the duration of $t_\mathrm{dyn}$ and only interact with other particles through gravity. In practice, the most of the star formation threshold densities in our simulations span the range between $10^{2}$ cm$^{-3}$ and $2\times 10^{5}$ cm$^{-3}$ (see also \citealt{2022MNRAS.509.5938H}), corresponding to $t_\mathrm{dyn}$ in the range from $\sim 0.04$ Myr to $\sim 3$ Myr.

The mass of the inert reservoir particles that exceed the assigned $t_\mathrm{dyn}$ is then sampled randomly into individual stars along the Kroupa IMF \citep{2001MNRAS.322..231K} with an allowed range between $0.08$--$500$ \mdot. When the total sampled mass for one particle exceeds the mass of the particle (typically $\sim4$ \mdot) by more than a pre-defined tolerance of $\pm0.008$ \mdot{} (10\% of the minimum mass), local conservation of mass is enforced by searching for other inert reservoir particles within a radius of 1 pc. The radius of  1 pc corresponds to the Jeans-length at the typical star formation densities in  our present simulations. Then, if enough nearby reservoir particles are present, one neighbour at a time transfers mass to the particle being sampled as long as the mass excess is satisfied. Metals are transferred according to the fraction of mass being transferred by the donor particle. A neighbour may donate its entire mass, and in this case the neighbour particle is immediately removed from the simulation. 

The sampling of the reservoir particles proceeds in a random order, one particle at a time, over all of the reservoir particles that exceed their $t_\mathrm{dyn}$. When mass is borrowed from neighbours, the properties of the neighbours are updated immediately to ensure mass is always conserved. Particles in denser regions will initiate the sampling process due to the shortest $t_\mathrm{dyn}$, therefore they will also have the largest number of neighbouring reservoir particles and the highest chance of spawning massive stars. If there is not enough mass to be transferred within the search radius of a particle, the particle being sampled is skipped on this time step. A particle with failed IMF sampling can, however, still donate mass to other nearby reservoir particles when it falls within their individual search radii. In the case when the mass of a particle is not successfully sampled or transferred entirely to other nearby particles, the process is repeated for that reservoir particle on the next time step. A similar approach was recently adopted by \citet{2021PASJ...73.1036H} who introduced a searching radius approach for local mass conservation. Their approach uses the surrounding gas distribution as the reservoir, and a stricter favoured search radius of 0.2 pc.

At the gas densities where stars form in our simulations, a 1 pc sphere of uniform density would in theory encompass a reservoir of $4\pi\rho/3 (1\,\mathrm{pc})^3=10$--$10^4$ \mdot. The highest densities in our simulation could therefore in the best case scenario allow for stars with masses of hundreds of \mdot{} to form. We let the routine sample up to an upper stellar limit of $500$ \mdot{}, however the typically low star formation rates (SFRs) in our isolated low-metallicity dwarf galaxy models never allow the reservoir to grow large enough to realize the highest stellar masses beyond \mbox{$\sim 150$ \mdot}. In practice, for the majority of the $\sim4$ \mdot{} gas particles we spawn up to tens of new low mass stars per particle.

The newly sampled individual stars are spatially distributed according to a Gaussian distribution around the parent particle with standard deviation equal to the gravitational softening length of \mbox{0.1 pc}, with the central \mbox{$10^{-3}$ pc} and outer $>1$ pc excluded. The particle velocities are also randomly shifted by a Gaussian with standard deviation of \mbox{$0.1$ km s$^{-1}$}, with \mbox{$<10^{-3}$ km s$^{-1}$} and \mbox{$>1$ km s$^{-1}$} excluded, selected to be slightly smaller than the typical velocity dispersion values of molecular clouds \citep{1987ApJ...319..730S} and cold gas clouds in our dwarf galaxy simulations. A similar random velocity spread was employed by \citet{2023MNRAS.521.2196A} who sampled single stars down to \mbox{2 \mdot}. Once new stars have been successfully sampled, they may begin releasing radiation and, in the case of massive stars, stellar winds according to the individually assigned stellar masses. 

\begin{figure}
\includegraphics[width=0.9\columnwidth]{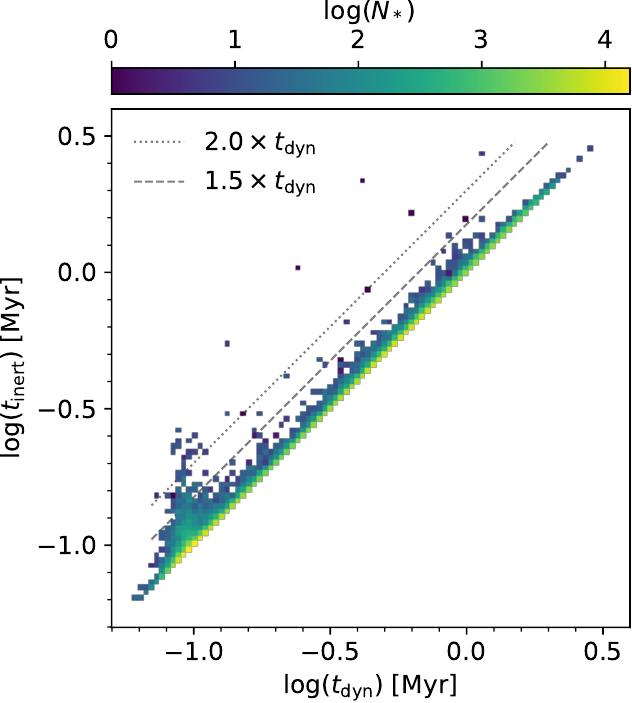}
\caption{The dynamical time $t_\mathrm{dyn}$ at the star-formation threshold density versus the total time $t_\mathrm{inert}$ the reservoir particles spent before spawning new stars in the compact dwarf run with winds. The inert time is a combination of $t_\mathrm{dyn}$ and the time it takes to succeed in the IMF sampling. All individual stars formed in the simulation are shown, the number of stars per pixel is indicated with the colour, and the grey lines indicate values of $1.5 \times t_\mathrm{dyn}$ and $2 \times t_\mathrm{dyn}$. The majority of the stars form within $1.5\times t_\mathrm{dyn}$. \label{fig:tdyn}}
\end{figure}

Fig. \ref{fig:tdyn} shows the realized delay time each reservoir particle first stays inert before finalizing the IMF sampling, as a function of the initial dynamical time of the star-forming gas in the compact dwarf galaxy model (see Section \ref{section:ICs}). The majority of the stars are sampled immediately after the dynamical time has passed, or after up to a couple dynamical times. In rare cases the process takes longer than twice the dynamical time, in total for $\sim200$ stars (0.05\%). 

\subsection{Stellar feedback}

\begin{figure*}
\includegraphics[width=\textwidth]{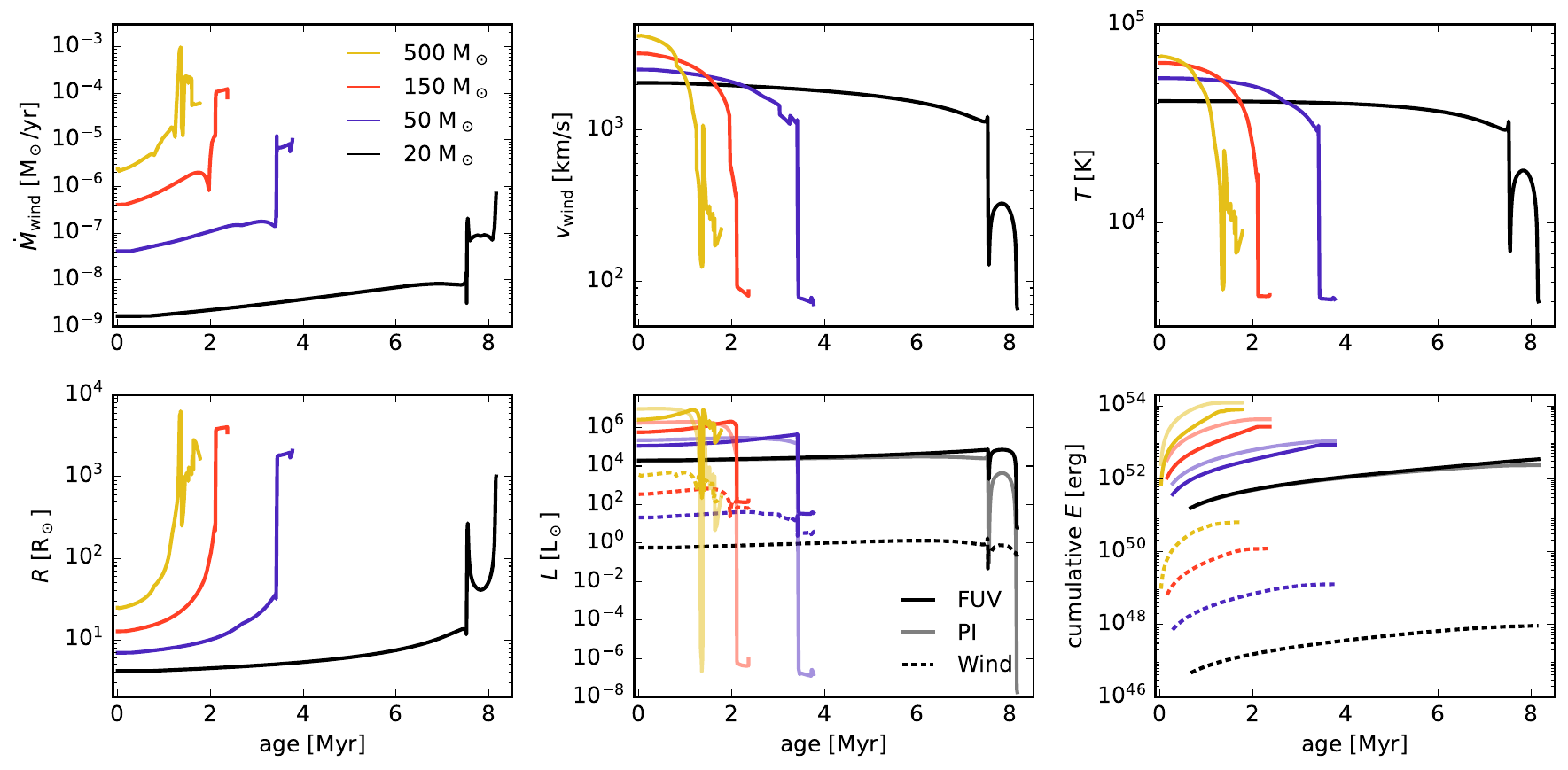}
\caption{The evolution of mass-loss rate ($\dot{M}_\mathrm{wind}$), wind velocity ($v_\mathrm{wind}$), surface temperature ($T$), stellar radius ($R$), radiative (FUV and photoionizing, PI) and mechanical luminosities and the cumulative energy in the wind and radiation components for four stars of 20 \mdot, 50 \mdot, 150 \mdot{} and 500 \mdot{} at $Z=0.00021$. The luminosity and integrated energy of the wind are shown with dashed lines. The cumulative energies in the last panel start at the first non-zero time step of each track. \label{fig:tracks}}
\end{figure*}

\subsubsection{Evolutionary tracks of massive stars}

For this study we implement the novel stellar models from Bonn Optimized Stellar Tracks (BoOST, \citealt{2022A&A...658A.125S}) that provide evolutionary tracks for stellar masses in the range of $9$--$500$ \mdot{} and metallicites from solar\footnote{We assume $Z\sim0.013$ for the solar value \citep{2009ARA&A..47..481A}.} down to the metallicity of the lowest metallicity dwarf galaxies. From the range of models, we select the slowly rotating low-metallicity IZw18 stellar tracks that represent a typical dwarf galaxy metallicity at \mbox{$Z=0.00021=0.016$ \zdot}. Here we assume a fixed metallicity for the pre-supernova evolution of the massive stars. The stars in our simulations can have a range of metallicities that we account for in the SN and AGB feedback prescription, and we will in future applications account for the metallicity-dependent pre-supernova evolution of the massive stars. The BoOST tracks consist of ten stellar models between $9$--\mbox{$500$ \mdot{}} at each value of metallicity, that have been further interpolated into a tighter grid of 1856 tracks spaced logarithmically by the initial mass. Each track has 608 time stamps along the evolution of the star, spaced according to the evolutionary phases of each stellar model. 

For this and our future studies the most important aspect of the BoOST tracks are the detailed surface abundances and the stellar wind mass-loss properties that are traced until the end of core helium burning.  We track in total 13 elements released via stellar feedback: H, He, N, C, O, Si, Al, Na, Mg, Fe, S, Ca and Ne. Out of these elements, the mass of Ca can only increase via SNe (Section \ref{section:supernovae}); Na and Al by SNe and the stellar winds of massive stars (Section \ref{section:stellar_winds}); and S by SNe and by winds of AGB stars (Section \ref{section:supernovae}). In the context of star clusters, the most important elements are the light elements Na, O, Al and Mg, that provide the most commonly observed abundance ratios in massive star clusters (see e.g. \citealt{2018ARA&A..56...83B, 2019A&ARv..27....8G} and the references therein). Fe, for instance, can be used to trace SN enrichment. Each element released via stellar winds of massive stars, AGB stars and SNe is followed separately for each particle, allowing us to decode which enrichment process contributed to the elemental abundances of each individual star and gas particle. 

\subsubsection{Winds of massive stars}\label{section:stellar_winds}

The BoOST tracks provide the surface abundances of the chemical species relevant for tracing the pre-SN metal enrichment around massive stars. Such detailed stellar tracks are crucial, for instance, for the modelling of the formation of multiple populations in globular clusters \citep{2019ApJ...871...20S} that are characterized by specific patterns in elemental abundance ratios, such as the Na--O and/or Mg--Al anticorrelation \citep{2018ARA&A..56...83B, 2019A&ARv..27....8G}. Our simulations follow element-by-element the continuous, evolving stellar wind released by each star more massive than $9$ \mdot{}\footnote{Note that in our model all stars more massive than 8 \mdot{} release FUV and photoionizing radiation and explode as SNe, but we only follow winds for stars more massive than $9$ \mdot{} due to the range of the BoOST tracks. The model can be supplemented with any set of stellar tracks.} individually. Here we track the stars until the end of the helium burning phase and neglect the latest phases of nuclear burning from carbon-burning onward, as these phases would last only up to a few kyr \citep{2013MNRAS.433.1114Y, 2013A&A...558A.103G}. We convert the evolving surface abundances and the total mass-loss rates into mass-loss rates of each individual element that are then released into the surrounding ISM at the wind velocity given by the stellar tracks. Representative examples of the wind mass-loss rates and wind velocities are shown in Fig. \ref{fig:tracks}.

Following similar approaches by \citet{2015ApJ...798...32N}, \citet{2017MNRAS.464.3536R}, \citet{2019MNRAS.483.3363H} and \citet{2022MNRAS.510.5592A} among others (see also discussion e.g. in \citealt{2018MNRAS.477.1578H} and \citealt{2022MNRAS.514..249C}), we distribute the wind into the ambient ISM using the \healpix{} algorithm \citep{2011ascl.soft07018G}. \healpix{} divides the $4\pi$ solid angle into equal sized pixels, where the pixel resolution can be increased by factors of four starting from 12 pixels. Throughout this work we use the base resolution of 12 pixels. In order to distribute feedback to approximately 100 gas particles, the code then searches within each 12 pixels for the $N_\mathrm{gas,pix}=8\pm 2$ nearest neighbouring gas particles. Each pixel receives an equal fraction of the wind mass $M_\mathrm{wind}$, and each $N_\mathrm{gas,pix}$ gas particles within one pixel receives an equal fraction, thus the received wind mass element per one gas particle is
\begin{equation}\label{eq:windmass}
    m_\mathrm{wind}=\frac{M_\mathrm{wind}}{12 N_\mathrm{gas,pix}}.
\end{equation}
The momentum of the wind is injected assuming an inelastic collision between the gas particles and the radially expanding wind, resulting in a boost in velocity given by \begin{equation}
    \Delta \textbf{\textit{v}}_\mathrm{gas} = \frac{m_\mathrm{wind}}{m_\mathrm{gas}+m_\mathrm{wind}}(\textbf{\textit{v}}_\mathrm{wind}-\textbf{\textit{v}}_\mathrm{gas})
\end{equation}
where $m_g$ is the mass of the gas particle and $\textbf{\textit{v}}_\mathrm{gas}$ and $\textbf{\textit{v}}_\mathrm{wind}$ are the velocity vectors of the gas particle and the wind element. The stellar wind is expelled radially at a velocity given by the respective BoOST track. The corresponding increase in internal energy of the wind-receiving gas particle is given by
\begin{equation}
    \Delta E_\mathrm{int} = \frac{1}{2}\frac{m_\mathrm{gas} m_\mathrm{wind}}{m_\mathrm{gas}+m_\mathrm{wind}}(\textbf{\textit{v}}_\mathrm{wind}-\textbf{\textit{v}}_\mathrm{gas})^2 .
\end{equation}

\begin{figure*}
\includegraphics[width=\textwidth]{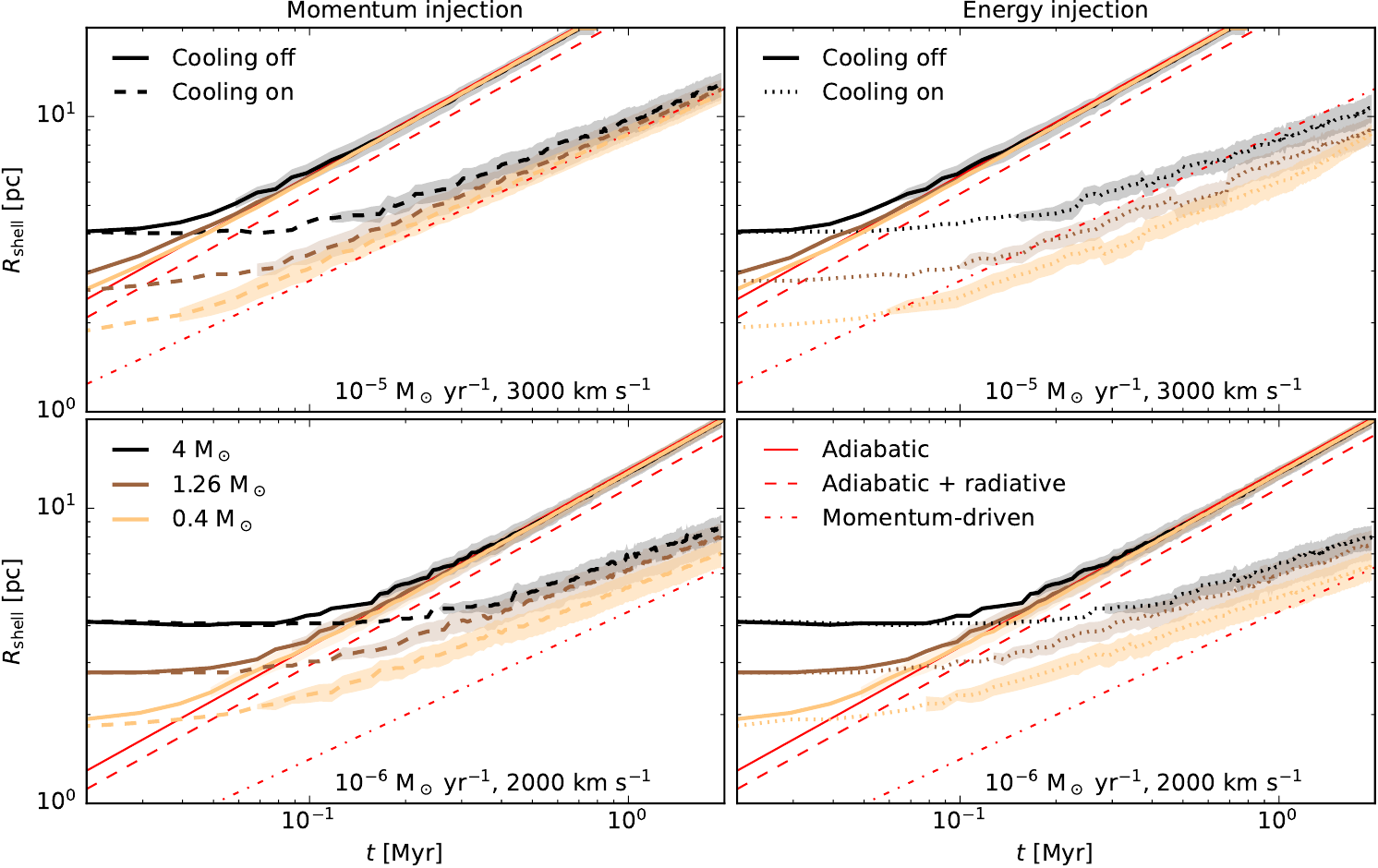}
\caption{The evolution of the wind shell radius $R_\mathrm{shell}$, defined as the peak of the radially averaged gas density profile, around a massive star interacting with an ambient medium with $\rho_0=2\times 10^{-22}$ g cm$^{-3}$, $T=20$ K and solar metallicity. On the left the kinetic energy of the wind is injected by adding momentum to the receiving gas particles, while on the right we inject the kinetic energy as thermal energy into the gas particles. The thick solid lines show the evolution of the wind bubble with cooling switched off, analogous to the adiabatic expansion in Eq. \eqref{eq:energy} with $R_\mathrm{shell}\propto t^{3/5}$ that is shown with the red solid line. The dashed (left) and dotted (right) coloured lines use the full code implementation including low temperature cooling. The red dashed lines show Eq. \eqref{eq:energy_rad} that includes radiative cooling, and the red dot-dashed lines show Eq. \eqref{eq:momentum} with $R_\mathrm{shell}\propto t^{1/2}$ which demonstrate momentum-conserving expansion. Three mass resolutions of $4$ \mdot{} (black), $1.26$ \mdot{} (brown) and $0.4$ \mdot{} (yellow) for the ambient medium are shown in each panel. All panels assume constant mass-loss rates, with $\dot{M}_\mathrm{wind}=10^{-5}$ \mdot{} yr$^{-1}$ and $v_\mathrm{wind}=3000$ km s$^{-1}$ in the top row and \mbox{$\dot{M}_\mathrm{wind}=10^{-6}$ \mdot{} yr$^{-1}$} and $v_\mathrm{wind}=2000$ km s$^{-1}$ in the bottom row. The shaded regions show the standard deviation of radii for gas particles with density $\rho>1.2\rho_0$ once such particles exist. \label{fig:Rshell}}
\end{figure*}

We test the wind model by placing a very massive star that releases a constant stellar wind with a mass-loss rate $\dot{M}_\mathrm{wind}$ and a wind velocity $v_\mathrm{wind}$ in a box of ambient gas with a side length of $\sim50$ pc and initial density $\rho_0=2 \times 10^{-22}$ g cm$^{-3}$, temperature \mbox{$T=20$ K} and solar metallicity. This metallicity is chosen as an upper limit and presents the most challenging conditions in terms of cooling, while our galaxy simulations are performed at a lower metallicity. For $\dot{M}_\mathrm{wind}$ we selected \mbox{$10^{-6}$ \mdot{} yr$^{-1}$} and \mbox{$10^{-5}$ \mdot{} yr$^{-1}$} and for $v_\mathrm{wind}$ values of $2000$ km s$^{-1}$ and $3000$ km s$^{-1}$, respectively, that are typical values for the stellar winds of very massive stars. Radiation is not included for these tests as photoionization in particular may dominate the energy-balance in the vicinity of the star especially in the case of a cold neutral medium \citep{2018MNRAS.478.4799H}. We conducted two sets of runs, without and with cooling, and compute the evolution of the radius of the expanding wind shell ($R_\mathrm{shell}$) from the maximum of the radially averaged density profile. For the ambient gas mass resolution we test three values separated by 0.5 dex; $4$ \mdot{} (the fiducial resolution of the galaxy simulations), $1.26$ \mdot{} and $0.4$ \mdot.
The tests were additionally run using the direct injection of the wind luminosity as thermal energy instead of imparting the energy as an update in the momentum, to verify that the wind is not resolved enough to generate the wind momentum as is the case for SNe (see e.g. \citealt{2016MNRAS.458.3528H} and \citealt{2020MNRAS.495.1035S}). 

The hydrodynamical models are compared in Fig. \ref{fig:Rshell} to the analytic solution for the radius $R_\mathrm{shell, E}$ of a thin shell expanding adiabatically \citep{1977ApJ...218..377W, 2001PASP..113..677C} with time $t$ as
\begin{equation}\label{eq:energy}
    R_\mathrm{shell,E}=0.88\left(\frac{L_\mathrm{wind}}{\rho_0}\right)^{1/5}t^{3/5},
\end{equation}
where $L_\mathrm{wind}=0.5\dot{M}_\mathrm{wind}v_\mathrm{wind}^2$ is the mechanical luminosity of the stellar wind. Eq. \eqref{eq:energy} corresponds to the first phase of wind bubble expansion. If the shell is allowed to cool radiatively, corresponding to the next stage of bubble evolution, the expansion slows down slightly and the solution becomes \citep{1977ApJ...218..377W}
\begin{equation}\label{eq:energy_rad}
    R_\mathrm{shell,ER}=0.76\left(\frac{L_\mathrm{wind}}{\rho_0}\right)^{1/5}t^{3/5}.
\end{equation}
Both of the solutions above with $R\propto t^{3/5}$ describe how the adiabatic expansion should result in a momentum $p\propto t^{7/5}$ (see Fig. \ref{fig:momentum} in Appendix \ref{appendix:A}), that is boosted compared to the wind momentum $p\propto t$ alone (see e.g. discussion in \citealt{2021MNRAS.508.1768P}). A third solution assumes mixing between the ambient low-temperature medium and the wind shell boundary, which results in efficient cooling and a ram-pressure driven momentum-conserving expansion of the wind shell radius $R_\mathrm{shell,M}$ \citep{1975ApJ...198..575S, 2001PASP..113..677C} given by
\begin{equation}\label{eq:momentum}
    R_\mathrm{shell,M}=\left(\frac{3}{2\pi}\frac{\dot{M}_\mathrm{wind}v_\mathrm{wind}}{\rho_0}\right)^{1/4}t^{1/2}.
\end{equation}
With efficient cooling, the momentum boost from the adiabatic expansion is lost and the bubble evolves according to the wind momentum $p\propto t$. In a turbulent ambient medium, such as a galactic ISM, energy losses can be driven by both mixing and venting of the hot gas through less dense channels. \citet{2021ApJ...914...89L} showed that in a turbulent medium the wind shell will likewise evolve according to $R_\mathrm{shell}\propto t^{1/2}$. 

The wind is injected evenly into the $(8\pm2)\times 12$ neighbours in the \healpix{} pixels that extend similar to the SPH kernel in the initially uniform medium. The expansion of the simulated wind bubbles in Fig. \ref{fig:Rshell} begins with a head start in shell radius of a few pc compared to the analytic solutions from Eq. \eqref{eq:energy}--\eqref{eq:momentum}. The evolution of the wind bubble in the runs without cooling is otherwise identical to the analytic solution in Eq. \eqref{eq:energy}. In the runs that include cooling, the efficient cooling in the bubble is instead apparent. The PdV work in the bubble should result in an expansion according to Eq. \eqref{eq:energy_rad}, however neither of our implementations nor any of the mass resolutions can fully create the additional momentum due to cooling. Our fiducial momentum injection scheme performs better (as demonstrated e.g. by \citealt{2021MNRAS.508.1768P}) and results always in some additional momentum, but not enough to satisfy Eq. \eqref{eq:energy_rad}. The radial momentum in the bubble remains positive throughout the test in all cases, therefore all of the tests generate a reasonable bubble regardless of the energy losses. 

The gas mass resolution plays a minor role in the evolution of the wind shell, and the injection schemes differ very little in the low injection rate runs. In addition to the test runs shown in Fig. \ref{fig:Rshell} where solar metallicity was used, we also tested the \mbox{$\dot{M}_{w}=10^{-5}$ \mdot{} yr$^{-1}$} mass-loss models at \mbox{0.1 \zdot{}} metallicity. For the full model with cooling, lower metallicities reduce the cooling rate towards the behaviour of the adiabatic models (Eq. \ref{eq:energy} and  \ref{eq:energy_rad}). In practice, however, at \mbox{$Z=0.1$ \zdot{}} only the runs with thermal energy injection were affected and showed only a slightly reduced deceleration.

\citet{2014MNRAS.442..694D}, \citet{2015ApJ...798...32N}, and \citet{2022MNRAS.510.5592A} have demonstrated qualitatively similar bubble evolution for similar injection methods as ours. When considering the conclusions made by such studies on the weak effect of winds, we have to keep in mind that they do not fully account for the momentum input of the stellar wind. Such is the case for our simulations as well, as demonstrated in Figures \ref{fig:Rshell} and \ref{fig:momentum}. The wind bubbles evolve more as one would expect for the turbulent ISM \citep{2021ApJ...914...89L, 2021ApJ...922L...3L}. Stellar wind injection studies that better account for the energy input, such as \citet{2018MNRAS.478.4799H}, have shown that winds can overcome photoionizing radiation in certain conditions (e.g. in warm low-density medium). However, in more realistic cloud-scale studies by e.g. \citet{2021MNRAS.506.3239G}, as long as radiation input is included, the addition of well resolved winds does not affect the realized star formation or the IMF \citep{2022MNRAS.515..167G}.

\begin{figure*}
\includegraphics[width=\textwidth]{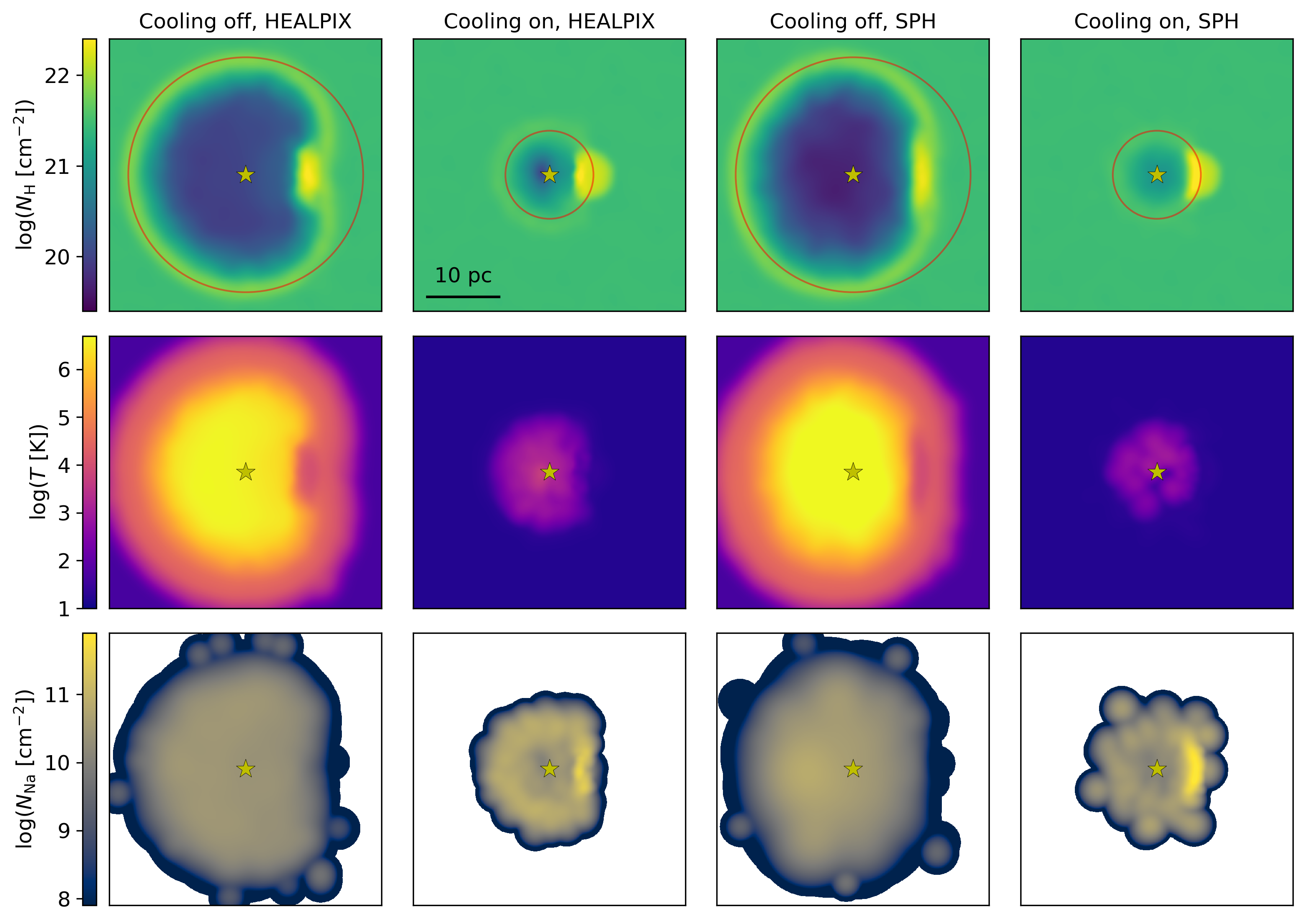}
\caption{The hydrogen column density (top), temperature (middle) and sodium column density which illustrates the metal enrichment (bottom) of gas around a massive star approximated with constant $\dot{M}_\mathrm{wind}=10^{-5}$ \mdot{} yr$^{-1}$ and $v_\mathrm{wind}=3000$ km s$^{-1}$, interacting with a 1000 \mdot{} gas cloud with density \mbox{$\rho_0=2\times 10^{-21}$ g cm$^{-3}$}. The two left columns show runs where wind injection is performed using the \healpix{} implementation while the two right columns use SPH weighting. The snapshots are taken after 0.5 Myr of constant stellar wind. The ambient medium has $\rho_0=2\times 10^{-22}$ g cm$^{-3}$, $T=20$ K and solar metallicity. The slices span 38 pc across and the thickness of the slices is equal to two SPH smoothing lengths of the ambient gas ($\pm 3.3$ pc). The red circles show the analytic solution from Eq. \eqref{eq:energy} (top) and Eq. \eqref{eq:momentum} (bottom) at equivalent time. The data in each panel have been smoothed using an SPH kernel and white pixels have no data. \label{fig:blob}}
\end{figure*}

Finally, we demonstrate the advantage of using \healpix{} to distribute the wind feedback compared to SPH weighting by showing in Fig. \ref{fig:blob} four additional runs at solar metallicity and 4 \mdot{} resolution. Fig. \ref{fig:blob} shows runs without and with cooling using both \healpix{} and SPH kernel weighting. A spherical cloud with a ten times higher density of \mbox{$\rho_0=2 \times 10^{-21}$ g cm$^{-3}$} and radius of \mbox{3.5 pc} is placed so that the wind-releasing star is at a distance of \mbox{2 pc} from the outer radius of the cloud. The snapshots in Fig. \ref{fig:blob} are taken after \mbox{0.5 Myr} of wind with $\dot{M}_\mathrm{wind}=10^{-5}$ \mdot{} yr$^{-1}$ blowing on the cloud. From top to bottom, the panels show hydrogen column density, gas temperature and sodium column density, assuming the wind material has the evolving surface composition of a 150 \mdot{} star. The sodium column density traces the metals injected into the gas. We do not include metal diffusion in our current model (see e.g. \citealt{2013MNRAS.434.3142A}, \citealt{2019MNRAS.483.3363H}) to have an unambiguous definition for enriched gas, therefore the outer regions of the wind front show the SPH kernel smoothed surroundings of single enriched gas particles. The analytic solutions from Eq. \eqref{eq:energy} and Eq. \eqref{eq:momentum} at the corresponding time are indicated as well. 

In the \healpix{} runs, the wind bubble expands against and around the dense cloud which remains intact and colder than the expanding hot wind bubble. The wind energy is distributed fairly evenly across the $4\pi$ solid angle around the star (Eq. \ref{eq:windmass}), of which the cloud fills only a small fraction. In the runs with SPH weighting, the distance based weighting of the stellar wind injection instead favors the dense nearby cloud. The cloud ends up receiving a relatively larger fraction of the feedback, most clearly evident in the sodium distribution of the cooling run in the bottom right panel. In the run with cooling (right column) the wind bubble expands slowly, therefore the momentum and corresponding internal energy keep being injected preferentially into the cold dense cloud. The injected internal energy is radiated even more efficiently than in the corresponding \healpix{} run, and the SPH run struggles to maintain a hot spherical wind bubble.

\subsubsection{Radiation}
The spatially and temporally evolving interstellar radiation field drives the photoelectric heating rate according to the attenuated flux of all the individual stars more massive than 0.8 \mdot{} (for details see \citealt{2016MNRAS.458.3528H, 2017MNRAS.471.2151H}). Briefly, each gas particle is affected by the local far-ultraviolet (FUV, \mbox{$6$--$13.6$ eV}) flux of stars within a distance of 50 pc. The energy density is attenuated according to the optical depth, computed using \treecol{} \citep{2012MNRAS.420..745C} along 12 \healpix{} pixels around each gas particle. For gas chemistry and attenuation by interstellar dust, we assume a constant dust-to-gas mass fraction of $0.1Z=2.1\times 10^{-5}$ following the observed linear dependence between the gas metallicity and the dust-to-gas ratio \citep{2014A&A...563A..31R}\footnote{Note however that at low metallicity the dust-to-gas ratio may have even lower values, see \citealt{2014A&A...563A..31R}.}. Additionally, we model the photoionized regions around all stars more massive than \mbox{$8$ \mdot{}} using a Str\"omgren-type approximation given by the rate of ionizing photons \citep{2012MNRAS.421.3488H, 2017MNRAS.471.2151H}. All gas particles flagged as photoionized (see Section 2.5.2 in \citealt{2017MNRAS.471.2151H}) are kept at a temperature of $10^4$ K.

For all stars above \mbox{0.8 \mdot{}} we integrate the FUV fluxes using the effective temperatures from the low-metallicity Geneva stellar tracks at $Z=0.0004\sim 0.02$ \zdot \citep{2019A&A...627A..24G} and the \textsc{BaSeL} spectral library at corresponding to $Z\sim 0.01$ \zdot{} \citep{1997A&AS..125..229L, 1998A&AS..130...65L, 2002A&A...381..524W}. The low metallicity Geneva tracks lack values for stars below $1.7$ \mdot{}, thus we approximate the fluxes for \mbox{$0.8$--$1.7$ \mdot{}} stars by scaling up the fluxes in equivalent stellar mass range at $Z=0.002\sim 0.1$ \zdot{} from \citet{2013A&A...558A.103G}. We also integrate the photoionizing ($>13.6$ eV) radiation for 8--9 \mdot{} stars using the same procedure. For the wind-releasing stars for which we use the BoOST tracks ($>9$ \mdot) we instead integrate the black body spectrum and apply an opacity correction to account for the optical depth caused by electron scattering in the wind. Following \citet{1989A&A...210...93L} and  \citet[Chapter 4.5.1]{2016PhDT.......375S} we correct the radii of the stars as
\begin{equation}
    R_\mathrm{eff} = R + \frac{3}{2}\frac{\kappa \dot{M}_\mathrm{wind}}{4 \pi v_\infty}
\end{equation}
where $\kappa=\sigma (1+X)$ is the electron scattering opacity with $\sigma$ being the Thomson scattering cross-section and $X$ the hydrogen mass fraction, and $v_\infty =\sqrt{Gm_*/R}$ is the terminal wind velocity where $m_*$ is the mass of the star. As an example, the correction to $R$ for the models shown in Fig. \ref{fig:tracks} range from less than a percent in the lowest mass models and at times reach almost $50$\% in the 500 \mdot{} model. Using the Stefan-Boltzmann law and assuming $L=L_\mathrm{eff}$, the temperature of the star is then also corrected by
\begin{equation}
    T_\mathrm{eff} = T\sqrt{\frac{R}{R_\mathrm{eff}}}.
\end{equation}
Because $T_\mathrm{eff}$ depends on the square root of the true and corrected radius, the correction to $T$ in the models shown in Fig. \ref{fig:tracks} is typically small and up to 17\% for a short period of time in the highest mass model.

The two last panels in Fig. \ref{fig:tracks} show the opacity corrected values of FUV and photoionizing fluxes and their cumulative contribution compared to the wind luminosity for the four representative stellar masses. The photoionizing flux dominates the energy budget in the early stages of the evolution of the very massive stars, and the integrated mechanical luminosity is at least three orders of magnitude lower than either of the integrated radiative luminosities (FUV and photoionizing) even in the case of the most massive star of \mbox{500 \mdot}.

\subsubsection{Supernovae and AGB stars}\label{section:supernovae}
We assume that stars with initial masses between 8 \mdot{} and 50 \mdot{} explode as core-collapse SNe once they reach the end of their lifes. The evolution of the massive stars is followed until the end of the core helium burning phase as given by the BoOST tracks. We interpolate the metallicity-dependent SN yields logarithmically for the initial stellar mass range of 13--35 \mdot{} from \citet{2004ApJ...608..405C} and extrapolate the values outside of the range by scaling the ejecta mass down (below 13 \mdot) or up (above 35 \mdot) in logarithmic space. At the gas mass resolution of \mbox{$\sim 4$ \mdot{}} the evolution of SN remnants is resolved via injection of thermal energy \citep{2016MNRAS.458.3528H} in the majority of cases \citep{2022MNRAS.509.5938H} even in extreme star-forming environments \citep{2020ApJ...891....2L} because of the dispersing effect of early stellar feedback. Core-collapse SNe in our model release \mbox{$10^{51}$ erg} of energy into the surrounding ISM comprised of $(8\pm 2)\times 12$ gas particles, distributed using the \healpix{} algorithm similarly to the wind material \citep{2019MNRAS.483.3363H}. We note, that the explosion energy has a scatter in a wide range between $7\times 10^{49}$ erg and $2\times 10^{51}$ erg \citep{2016ApJ...818..124E,2020MNRAS.496.2039S} and will account for the energy spectrum and explodability in future applications.

The upper mass limit of core-collapse SNe depends on metallicity and is of the order of 40 \mdot{} at low metallicity \citep{2003ApJ...591..288H}. We assume a conservative limit of $50$ \mdot, wherein stars with higher initial masses collapse directly into black holes. An exception are stars that have final helium core masses between 65 and 133 \mdot{} that completely disrupt as pair-instability SNe (PISNe, \citealt{2002ApJ...567..532H}). In the BoOST tracks, such final helium core masses result from stars with initial masses between 107.2 \mdot{} and 203.4 \mdot. We have implemented the PISN ejecta and explosion energies ($5\times10^{51}$--$9\times10^{52}$ erg) of the zero-metallicity models from \citet{2002ApJ...567..532H}\footnote{\url{https://2sn.org/DATA/HW01/}} according to the mass of the helium core at the end of core helium burning given by the BoOST tracks. The zero-metallicity helium core models are used since differences to yields from models at non-zero metallicity have been shown to be small \citep{2014A&A...566A.146K, 2017ApJ...846..100G}. We supplement the elements released by the helium core with the rest of the mass of the star left over at the end of helium core burning. For the remaining mass we assume the final elemental abundances at the surface of the star according to the BoOST tracks. Otherwise the PISN ejection is performed similar to the core collapse SNe. The isolated dwarf galaxy runs presented here very rarely form stars with masses beyond $\sim100$ \mdot{}, thus the PISN implementation will be more relevant for future applications with more intense star formation.

Finally, AGB stars ($0.5$--$8$ \mdot) that in our model release a single burst of winds at the end of their lives are also included. AGB stars release metals according to metallicity dependent yields from \citet{2010MNRAS.403.1413K}\footnote{We do not include electron-capture SNe in the narrow range of $\pm 0.2$ \mdot{} around $\sim8.8$ \mdot{} even though their nucleosynthetic yields differ from iron
core-collapse explosion \citep{2013ApJ...774L...6W}.}. However, in the present study we concentrate on the enrichment by SNe and winds of massive stars and refer to the latter simply as stellar winds.

\begin{figure*}
\includegraphics[width=\textwidth]{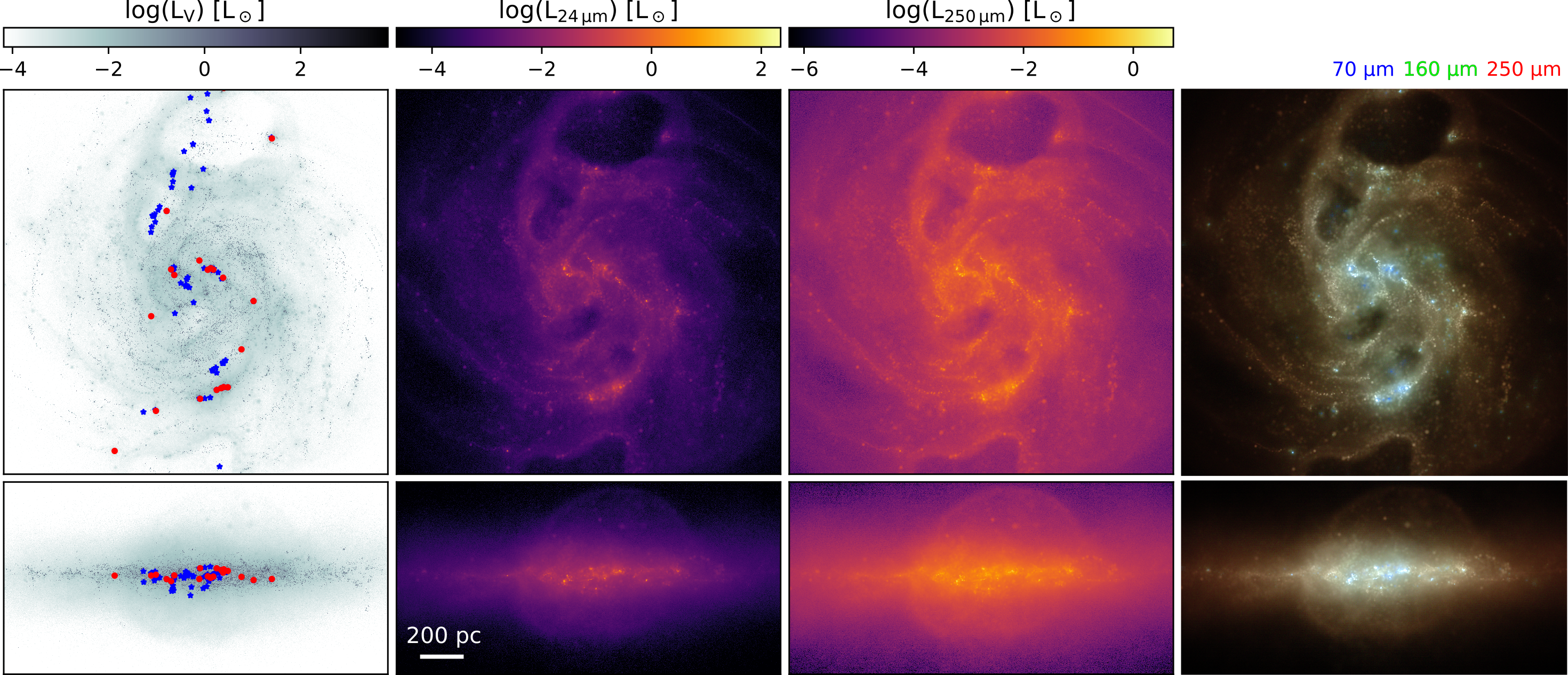}
\caption{The V-band, 24 $\mu$m and 250 $\mu$m emission of the compact dwarf galaxy after 500 Myr of star formation in face-on and edge-on projections. The V-band image highlights locations of bright stars and star clusters as well as their scattered emission, while the 24 $\mu$m map shows hot and more localized dust emission compared to the less localised cooler dust emission in the 250 $\mu$m map. The maps span 2 kpc across. The symbols on the left indicate massive stars (>8 \mdot, blue) and young bound star clusters (<10 Myr, red). The rightmost column shows a composite image of 70 $\mu$m, 160 $\mu$m and $250\mu$m emission, where the young star clusters and massive stars are surrounded by hot dust emitting in shorter infrared wavelengths. The maps are best viewed on a computer screen. \label{fig:2dmaps}}
\end{figure*}

\subsection{Initial conditions and galaxy properties}\label{section:ICs}

The isolated gas-rich dwarf galaxy models with virial mass $M_\mathrm{vir}=2\times10^{10}$ \mdot{} and radius $r_\mathrm{vir}=44$ kpc employed in this study are similar to those utilized extensively in \citet{2016MNRAS.458.3528H}, \citet{2020ApJ...891....2L} and \citet{2022MNRAS.509.5938H}. We use two rotationally supported disk galaxy models, one with a compact and one with an extended gas disk. For both models the gas disk and the initial stellar disk have total masses of $4\times10^{7}$ \mdot{} and $2\times10^{7}$ \mdot{} with initial particle mass resolution of $\sim4$ \mdot. The initial, old stellar disks are passive in the sense that their stars do not contribute to stellar feedback. The rest of the virial mass is dark matter with particle mass resolution of $6.8\times10^3$ \mdot. The compact and extended dwarf galaxy models have gas disk scale lengths of $0.73$ and $1.46$ kpc, respectively. The stellar disk scale lengths are $1.46$ kpc and the disk scale heights are $0.35$ kpc for both models. For initial metallicity we set \mbox{$0.00021\sim 0.016$ \zdot{}} and the gravitational softening lengths of baryons and dark matter are 0.1 pc and 64 pc, respectively. Both of the galaxy models have been run using the full hydrodynamical implementation including stellar winds and with the winds switched off, therefore we present in total four simulation.

The dwarf galaxy models are run until star formation has proceeded for 500 Myr. In Fig. \ref{fig:2dmaps} we show the last snapshot of the fiducial compact dwarf galaxy model post-processed using the radiative transfer code \textsc{SKIRT} \citep{2020A&C....3100381C}. The methodology is the same as described in \citet{2022MNRAS.514.4560L}, except here we have placed the system at a closer distance of 3 Mpc, typical for star cluster studies in dwarf galaxies \citep{2012ApJ...751..100C, 2019MNRAS.484.4897C}. Fig. \ref{fig:2dmaps} shows broadband filtered images in V-band and \textit{Spitzer Space Telescope} \mbox{24 $\mu$m} and \mbox{$250$ $\mu$m} equivalent bands, and a composite of the \textit{Spitzer} \mbox{70 $\mu$m} and \mbox{160 $\mu$m} and \textit{Herschel Space Observatory} 250 $\mu$m equivalent bands\footnote{A higher resolution version of the figure can be found \href{https://wwwmpa.mpa-garching.mpg.de/~naab/griffin-project/observations.html}{here}.}. The images have an intrinsic resolution 0.031\arcsec, corresponding to the pixel scale of UV-visual bands of \textit{Hubble Space Telescope} (\textit{HST}) and \textit{James Webb Space Telescope} (\textit{JWST}). The maps have been smoothed with a corresponding point spread function of FWHM=2.1 pixels. The random noise is omitted here simply for visual purposes, as most of the galaxy emission is below the noise level of a few \mbox{$10^{-8}$ Jy} that is typical for 1 h long visual and near-infrared observations with \textit{HST} or \textit{JWST}, and definitely below the larger values of noise in longer wavelength observations. The brightest pixels in UV-visual bands are of the order of $4\times 10^{-6}$ Jy after the 2.1 pixel PSF is added. In the best case scenario, where the bright clusters would not be spread out too much at lowered resolution, the brightest star clusters could be detected at the 1 hr \textit{JWST} sensitivity ($1$--$2\times 10^{-8}$ Jy) out to tens of Mpc \footnote{\url{https://jwst-docs.stsci.edu/jwst-near-infrared-camera/nircam-performance/nircam-sensitivity}}. In reality the larger pixel scale at larger distances dilutes the pixelized luminosities, and especially the brightest point-like stars.

In the middle panels of Fig. \ref{fig:2dmaps}, the upper limit is capped to the brightest pixel in both images and the range spans the same order of magnitudes. The $250$ $\mu$m dust emission in the compact dwarf galaxy is less concentrated compared to the hot 24 $\mu$m dust emission. We show the currently active massive stars and bound star clusters younger than 10 Myr in the left hand panel. The blue regions in the composite image on the right indicate hot dust emission in the \mbox{24 $\mu$m} emission map, and correspond to the locations of the young star clusters and massive stars shown on the left.

\subsection{Detection of star clusters}\label{section:clusters}

We identify bound star clusters using the friends-of-friends algorithm with a linking length of 1 pc, combined with the \textsc{subfind} algorithm \citep{2001MNRAS.328..726S, 2009MNRAS.399..497D}. The minimum number of bound members for an object to be considered a cluster is set to 50, which for a mean stellar mass of \mbox{0.5 \mdot{}} and minimum mass of \mbox{0.08 \mdot{}} can result in clusters with total masses slightly below \mbox{25 \mdot}. We also derive some photometric properties for the cluster population using the projected emission maps produced with radiative transfer. For the photometry, we use the procedure of detection and aperture photometry described in \citet{2022MNRAS.514.4560L}. The pipeline follows \textit{HST} star cluster surveys at distances too far for resolved photometry and where aperture photometry is instead necessary. We use the B, V- and I-band maps to select clusters and perform aperture photometry with appropriate background and aperture corrections over the flux and stellar mass maps. The aperture radius is set to three pixels, corresponding to 1.35 pc. A \mbox{$3.2\times 10^{-8}$ Jy} gaussian random noise with 10\% standard deviation is added before starting the detection procedure.

\section{The star cluster population}\label{section:star_clusters}
\subsection{Cluster mass function}

We utilize the radiative transfer images produced in Section \ref{section:ICs} to compare the underlying bound star cluster population to the photometrically detected star cluster population in the compact dwarf galaxy model. Using the minimum number of stars required for a bound star cluster, we exclude photometrically detected apertures that include less than 50 stars. The number of stars per aperture is estimated using a number density map produced at the same spatial resolution as the emission maps. The photometric pipeline detects 99 unbound associations, chance projections and single stars in the low-mass end of the cluster mass function that are ecluded by the threshold limit of 50 stars. In observational surveys, single stars are excluded, for example, using the concentration of the light profile within the aperture \citep{1999AJ....118.1551W}.

\begin{figure}
\includegraphics[width=\columnwidth]{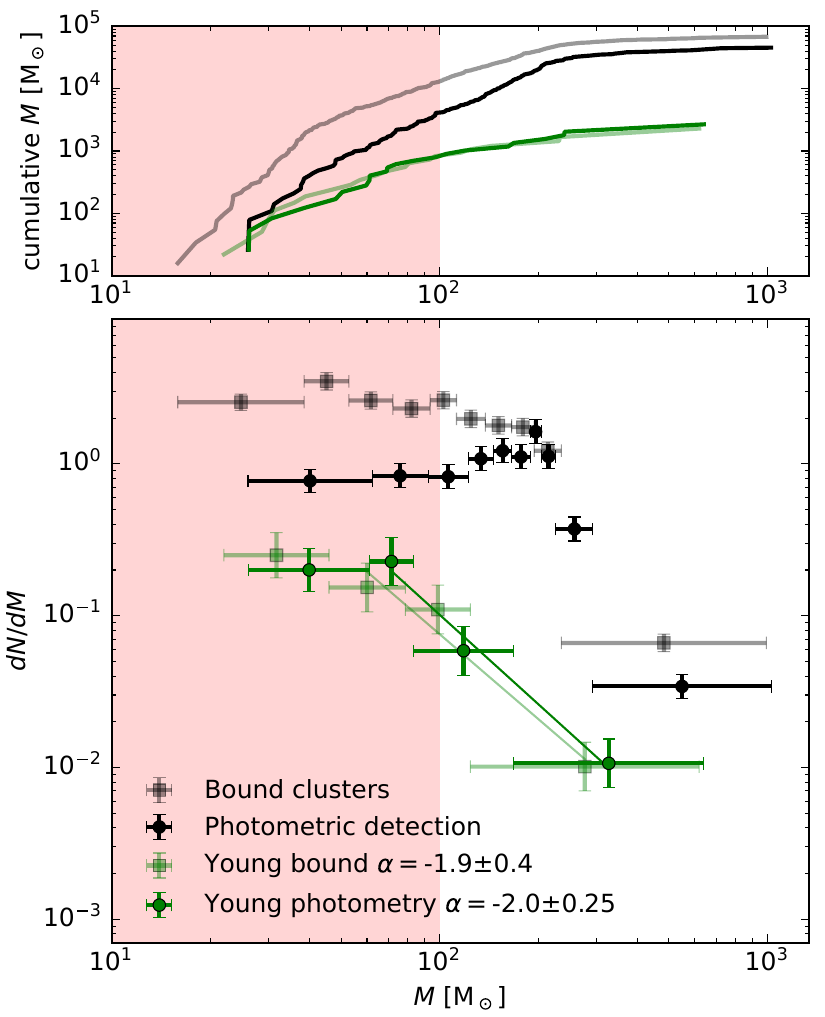}
\caption{A comparison of the masses of the photometrically detected clusters (black) and the bound star cluster population (gray) in the final snapshot of the compact dwarf galaxy run. Clusters younger than 10 Myr are shown in dark green for the photometry and light green for the bound clusters. Both data sets exclude clusters with less than 50 stars. The red shaded area indicates the limit of 100 \mdot{} that is the typical lower limit assumed in observed cluster surveys when the total cluster mass is integrated using a best-fit cluster mass function. The lines in the bottom panel show the best-fit mass function to the young cluster population, fit excluding the lowest mass bin, and the power-law indices with uncertainties are given in the legend. The top panel shows the cumulative mass in both of the data sets. The photometric pipeline recovers 67\% of the total bound cluster mass and overestimates the young cluster mass by 28\%. \label{fig:CMF}}
\end{figure}

We compare the photometrically detected and aperture integrated cluster masses to the bound star cluster masses in Fig. \ref{fig:CMF}. Clusters that have mass weighted mean stellar age younger than \mbox{10 Myr} are also shown separately. The top panel in Fig. \ref{fig:CMF} shows the cumulative mass of the two populations (all clusters and young clusters), using the two detection methods.  The cluster data have been binned with equal number of clusters per bin using ten bins in case of the whole population and as many bins as needed to satisfy five clusters per bin with the young clusters which are less numerous. The red area in the panels indicates the limit of 100 \mdot{} that is typically used as the lower limit of integration when an observed cluster mass function is translated into an estimate of the total cluster mass. 

The young star cluster population is consistent with the $-2$ slope as recovered by both of the detection methods, while the whole population shows a plateau in the low-mass end. We fit a power-law mass function to the young cluster population, excluding the smallest bin, and give the best-fit values in the legend of Fig. \ref{fig:CMF}. Changing the binning from 4 to 6 bins, the best-fit values range from $-1.7$ to $-2.0$. Most of the clusters with bound mass above \mbox{100 \mdot{}} are detected in the photometry while the plateau in the whole population below \mbox{100 \mdot{}} is underestimated. The photometric pipeline recovers in total approximately 67\% of the total bound cluster mass. On the other hand, most of the clusters that are younger than 10 Myr are detected. The total mass in young clusters is in fact overestimated compared to bound clusters by 28\%, which is similar to what we found for massive clusters in a starburst environment in \citet{2022MNRAS.514.4560L}. Young clusters tend to be grouped with other young clusters in the cluster-forming regions as can be seen in Fig. \ref{fig:2dmaps}, therefore source separation can be difficult when using fixed aperture sizes. More sophisticated methods, such as using an iterative aperture size as in \citet{2012ApJ...751..100C} would provide better accuracy in crowded cluster-forming regions. Our simulated cluster detection presents a best-case scenario in terms of image quality, since we only include random noise and a PSF and do not, for instance, consider confusion from background or foreground objects. We also do not limit the analysis to only round, concentrated bona fide clusters (as in Class 1 and 2 in LEGUS, \citealt{2017ApJ...841..131A}). The completeness limits in real observations are therefore typically somewhat higher in terms of cluster mass (e.g. \citealt{2017ApJ...839...78J, 2019MNRAS.484.4897C}).

\begin{figure}
\includegraphics[width=\columnwidth]{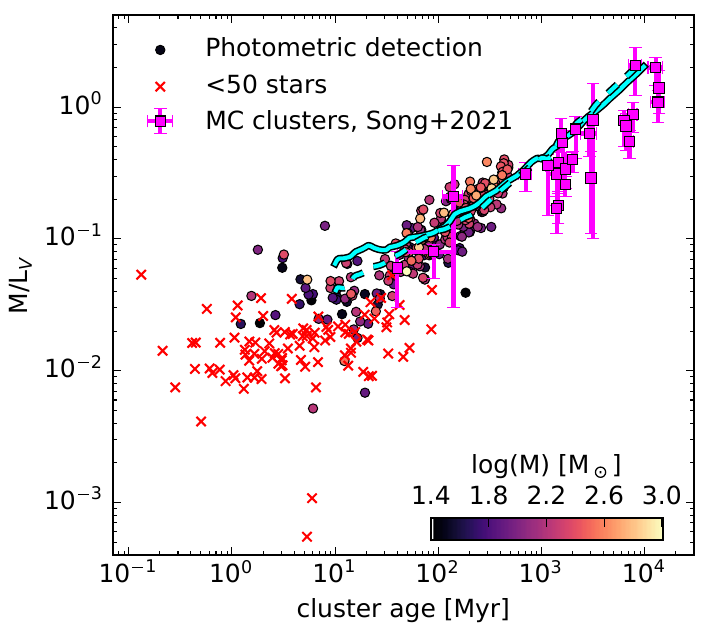}
\caption{The mass-to-light ratios of the photometrically detected clusters (circles) as a function of cluster age, colour coded by the cluster mass. The crosses show detections where the aperture includes less than 50 stars. The V-band luminosity and the total mass have been estimated using the photometric pipeline outlined in \citet{2022MNRAS.514.4560L}. Observed data for clusters more massive than $6\times 10^{3}$ \mdot{} in the Magellanic clouds reported in \citet{2021MNRAS.504.4160S} are shown with pink squares and errorbars. The cyan lines show the $M/L_\mathrm{V}$ values from the \citet{2003MNRAS.344.1000B} SSP library, assuming metallicities of $0.00021=0.016$ \zdot{} (solid) and $0.0021=0.16$ \zdot{} (dashed). \label{fig:MtoL}}
\end{figure}

\begin{figure*}
\includegraphics[width=\textwidth]{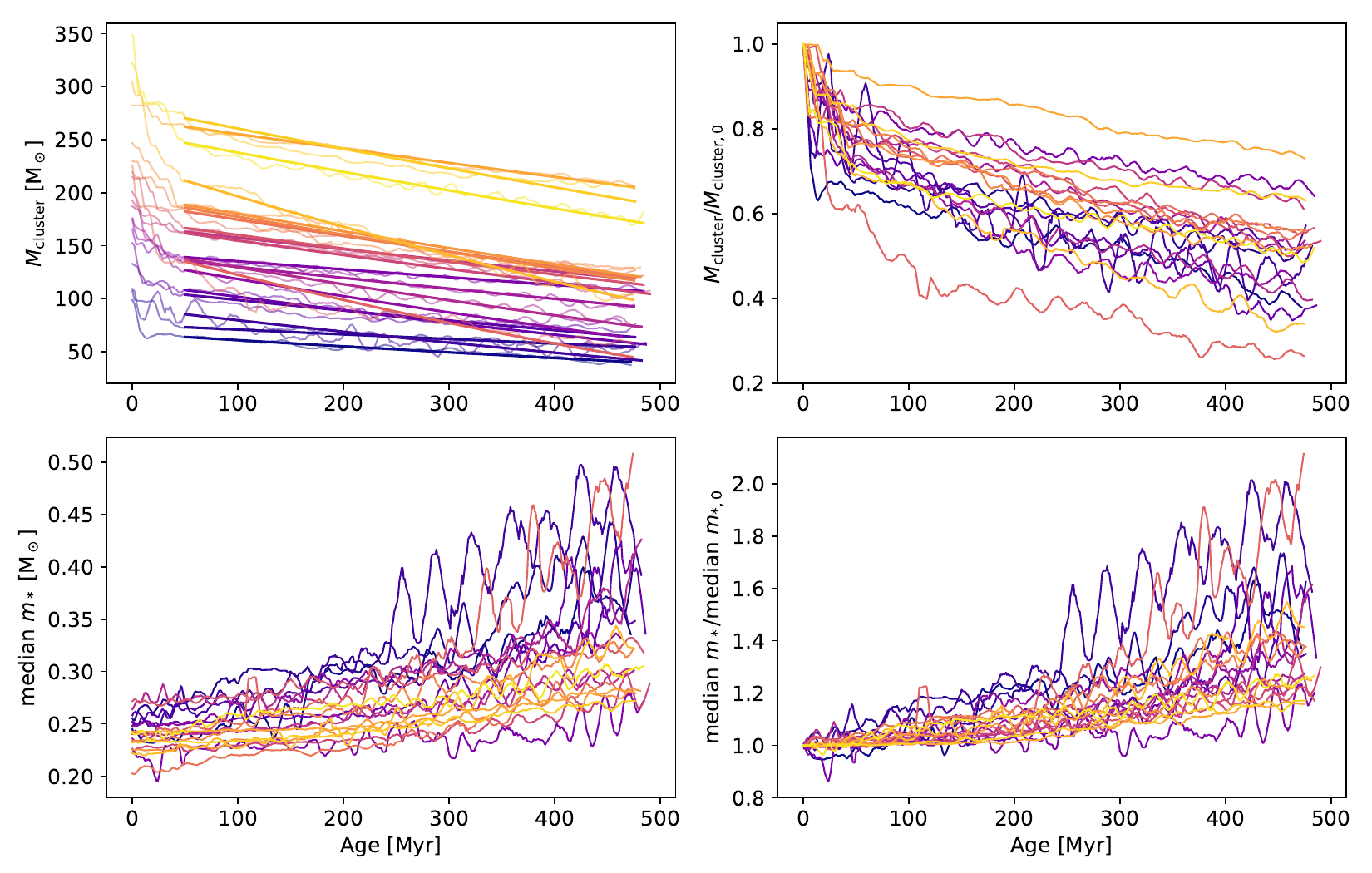}
\caption{Top: the mass evolution of the 20 most long lived clusters present in the final snapshot (500 Myr) of the compact dwarf galaxy simulation. The left and right hand panels show the total bound mass and the bound mass normalized to the initial mass. The clusters have been colour coded by the initial mass. The light lines on the background on the left show the evolution of each cluster, and the thick lines on top show a fit to a $dM/dt\propto M^{1-\gamma}$ mass-loss rate with a fixed slope of $\gamma=0.7$ (see text for details). The data have been smoothed over 5 Myr before age of 50 Myr and later over 10 Myr. Bottom: the median stellar mass (left) and the median mass normalized to the initial median mass (right) of bound stars in each of the clusters shown in the top row.} \label{fig:massloss}
\end{figure*}

\subsection{Mass-to-light ratio}

We compute the mass-to-light ratios ($M/L_\mathrm{V}$) of the photometrically detected clusters using the aperture integrated V-band luminosities and masses. Fig. \ref{fig:MtoL} shows the $M/L_\mathrm{V}$ as a function of mean cluster age. The data points are coloured by the cluster mass. We also show the detections that include less than 50 stars with separate symbols. All apertures with too few stars are located in the low $M/L_\mathrm{V}$ regime, in line with them being dominated by (a few) single bright stars.

The photometrically detected clusters are compared to a set of LMC and SMC clusters with various ages and, consequently, metallicities (between $[\mathrm{Fe}/\mathrm{H}]$ of $-2$ to $-0.3$, i.e. between $\sim 0.01$ \zdot{} and $\sim 0.5$ \zdot{}) from \citet{2021MNRAS.504.4160S}. The $M/L_\mathrm{V}$--age relation for simple stellar population (SSP) models from \citet{2003MNRAS.344.1000B} are also indicated for two metal mass fractions of 0.00021 and 0.0021, corresponding to our initial metallicity and a ten times higher value. The Magellanic clusters are more massive ($>6\times 10^{3}$ \mdot) and typically older than the clusters in our simulated galaxies. They populate a similar sequence, within errors, as the simulated clusters but at older ages and higher $M/L_\mathrm{V}$. The highest metallicity Magellanic clusters are predominantly located at lower $M/L_\mathrm{V}$ and young ages. For the simulated clusters, the $M/L_\mathrm{V}$--age relation is a result of stellar evolution and cluster mass-loss, as the stellar properties given to the radiative transfer were computed at fixed metallicity. In reality some of the clusters can have up to an order of magnitude enhanced metal content, even above $0.1$ \zdot. In the SSP model, a higher metallicity would lead to a lower $M/L_\mathrm{V}$ at young age ($<1$ Gyr). Additionally, we assume a black body approximation for the spectral energy distribution, which may introduce differences between our estimated $M/L_\mathrm{V}$ and the SSP models. In view of these uncertainties, the agreement between our radiative transfer post-processed star clusters and the SSP models is good.

\citet{2021MNRAS.504.4160S} showed that the Magellanic clusters were on average located below the $M/L_\mathrm{V}$--age plane estimated from SSP models, even when accounting for variations in metallicity. To explain this systematic shift, they discussed the work by \citet{2009A&A...502..817A} who studied the evolution of $M/L_\mathrm{V}$ using N-body models that account for dynamical mass-loss in addition to stellar evolution and metallicity. The evolving stellar mass function that becomes less populated by low-mass stars due to mass-loss in the tidal field, combined with stellar evolution, indeed results in lower $M/L_\mathrm{V}$ values compared to pure stellar evolution (i.e. SSP models) as a cluster ages. In our photometric sample, we can also see in Fig. \ref{fig:MtoL} a spread of values in $M/L_\mathrm{V}$ at a fixed age. Based on N-body models \citep{2002ApJ...576..899P} lower mass clusters could naively be expected to be further evolved at fixed age compared to higher mass clusters in terms of relaxation and mass-loss. In \citet{2009A&A...502..817A}, clusters which dissolve faster would deviate from the SSP model at a younger cluster age compared to clusters that survive for a longer time, corresponding e.g. to low and high-mass clusters evolving in an equivalent tidal field. In agreement with this, Fig. \ref{fig:MtoL} shows how it is the lower mass clusters (darker symbols) that fall preferentially below the sequence of most massive clusters (lightest symbols) at ages above a few tens of Myr. 

Another source for a spread in $M/L_\mathrm{V}$ could be the stochastic sampling of the stellar IMF. While pure stochasticity would lead to a spread both above and below a given fiducial SSP model in a cluster with a well sampled IMF, lower mass clusters are by construction less populated by brighter high-mass stars which should skew their resulting $M/L_\mathrm{V}$ toward higher values. However, the evolved clusters with low $M/L_\mathrm{V}$ values are in Fig. \ref{fig:MtoL} predominantly low mass clusters. They must have therefore either formed and retained an overabundance of massive stars, or evolved by removing low-mass stars in order to arrive at low values of $M/L_\mathrm{V}$. In the following we verify that clusters can both form with an excess of massive stars (Section \ref{section:imf}) and evolve by preferential removal of low-mass stars.

\subsection{Mass-loss}\label{section:massloss}

We check whether the evolution in $M/L_\mathrm{V}$ could be caused by dynamical mass-loss by investigating in Fig. \ref{fig:massloss} the evolution of the bound mass and the median stellar mass in clusters with respect to the cluster age. We show the properties of the 20 most long lived clusters present in the final snapshot of the compact dwarf galaxy. Clear trends in total mass and median mass can be seen in all of the clusters. The initial drop in total cluster mass is due to stellar evolution and early dynamical effects and the later evolution is caused by stars escaping due to the tidal field. The upward evolution in median mass in turn shows how the escaping stars have predominantly lower masses since the other driving factor, i.e. stellar evolution, should only drive the median mass downward. The small variations up and down in mass are due to the cluster detection routine where e.g. other nearby clusters can sometimes confuse the cluster boundaries. 

The thick lines in the top left panel of Fig. \ref{fig:massloss} show fits to the mass evolution using a pre-core collapse mass-loss rate estimate of the form 
\begin{equation}
    \frac{dM_\mathrm{cluster}}{dt}=-\frac{\mathrm{M_\odot}}{t_0} \left(\frac{M_\mathrm{cluster}}{\mathrm{M_\odot}}\right)^{1-\gamma}
\end{equation} where $t_0$ is a time scale parameter and $\gamma$ has typically values of 0.6--0.7 \citep{2003MNRAS.340..227B,2005A&A...441..117L}. Here we only fit the value of $t_0$ and fix the power-law slope to 0.7. The initial mass ($M_i$) is set to the mass after 50 Myr of evolution, when stellar evolution and early dynamical effects have diminished. The best fit values of the total dissolution time $t_\mathrm{dis}=t_0(M_i/M_\odot)^{\gamma}$ are between 0.9 Gyr and 3.1 Gyr. The mass-loss in our simulated clusters is mostly fit well by the mass-dependent mass-loss rate estimates, which are also found to fit observed clusters in nearby galaxies \citep{2003MNRAS.338..717B}. The observations include the SMC where clusters down to a few hundred \mdot{} were observed. The worst agreement is shown by the simulated cluster that lost more than 70\% of its mass. 

It is important to note how, even though the clusters have been sorted according to the initial mass, the curves at the end are not in the same order in either of the top panels of Fig. \ref{fig:massloss}. The local tidal field of a realistic dwarf galaxy affects each cluster differently, and the long-term evolution of the clusters cannot be determined unambiguously only by the initial mass. In N-body simulation \citep{2003MNRAS.340..227B}, the mass-loss rate depends for example on the initial density profile of the cluster.

In our set of 20 long-lived clusters, the clusters lost between 25\% and 70\% of their mass by the end of the simulation. Meanwhile, the median stellar mass increased at least 15\% and up to a factor of two. This long-term dynamical loss of low-mass stars can therefore explain some of the spread in the $M/L_\mathrm{V}$ values of the evolved clusters in Fig. \ref{fig:MtoL}. As a caveat, the clusters in our simulations are modelled using softened gravitational dynamics which underestimate the strength of close two-body encounters. Gravitational softening also precludes stellar multiples from hardening, hindering their regulating role as sinks and sources of energy in the evolving clusters. As a result, the clusters cannot undergo a dynamical core-collapse that may occur in later stages of the evolution of star clusters. The integrated mass-loss might therefore be underestimated and the cluster life-times overestimated. The total disruption times of the order of 1--3 Gyr for the simulated clusters are of similar order with the values estimated for a \mbox{$10^{4}$ \mdot{}} cluster in the SMC (8 Gyr) and the region around the Sun (1 Gyr) in \citet{2003MNRAS.338..717B}. Because the simulated clusters are smaller in mass, our total disruption times might be overestimated compared to the observed values. The weaker tidal field of the small mass dwarf galaxy can, however, also lead to a longer survival time compared to the more massive galaxies. To conclude, more accurate gravitational dynamics can be expected to mostly increase the spread towards even lower values of $M/L_\mathrm{V}$ in our "observed" cluster sample in Fig. \ref{fig:MtoL}. 

We do not address further the density profiles or the internal properties of the individual star clusters in the present study (see e.g. \citealt{2022MNRAS.510.5592A}). For the rest of the analysis we use clusters detected only with the \textsc{subfind} methodology.

\begin{figure}
\includegraphics[width=\columnwidth]{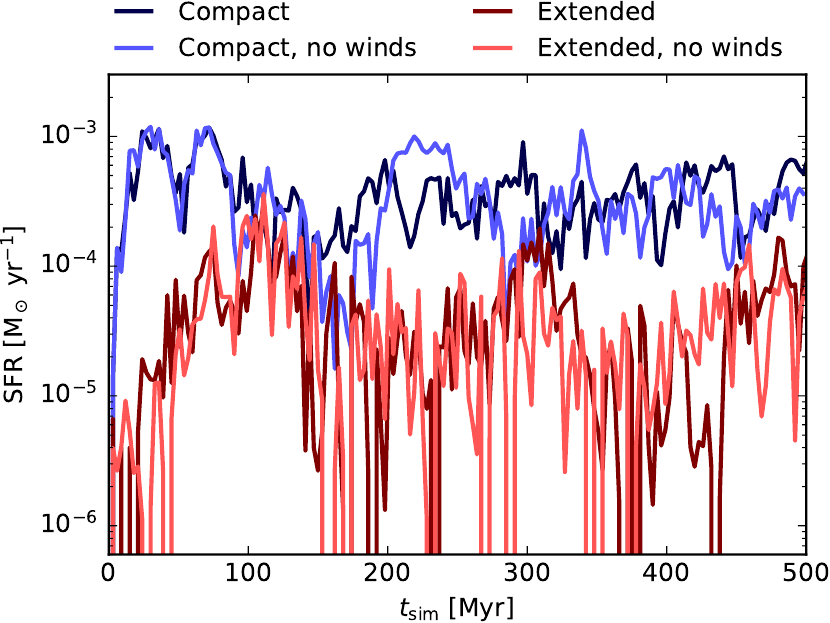}
\caption{The SFR in the isolated dwarf galaxy runs. The blue and red lines show the compact and the extended dwarf runs, respectively. The darker lines show the full models and the lighter lines show the runs with no winds. The SFR is computed as an average over the past 3 Myr, and the zero time stamp has been shifted to match when the first star forms in each run. \label{fig:sfr}}
\end{figure}

\section{Star formation}\label{section:star_formation}

In Fig. \ref{fig:sfr} we show the global integrated SFR in each of the isolated dwarf galaxy simulations with and without stellar wind enrichment. The SFR is an average over the past 3 Myr and the time axis has been shifted by the time it takes for the first stars to form in each simulation, i.e. $t_\mathrm{sim}=t-t_\mathrm{SFR}$. Due to the inefficient cooling at low metallicity, the compact dwarf spends the first 110 Myr without any star formation, whereas the extended dwarf evolves for 445 Myr before forming the first stars. For comparison, an otherwise equivalent compact dwarf model at 0.1 \zdot{} takes only 65 Myr to start forming stars. The compact dwarf has an average SFR of a few $10^{-4}$ \mdot{} yr$^{-1}$ while the extended dwarf forms stars at a ten times lower rate. The extended dwarf shows a somewhat bursty SFR evolution, where star formation is occasionally completely shut off for a period of a few Myr at a time. The runs with and without winds show very similar evolution, indicating that the stellar winds do not have a significant effect on the integrated star formation in either of the galaxy models as have been found in previous studies (e.g. \citealt{2012MNRAS.421.3488H, 2013ApJ...770...25A, 2014MNRAS.442..694D, 2018MNRAS.478.4799H, 2022MNRAS.510.5592A}). As noted in Section \ref{section:stellar_winds} and Appendix \ref{appendix:A}, the implemented method for stellar winds may underestimate the total momentum generated by the winds. In the turbulent galactic ISM, the mixing and venting of the heated gas may result in a lowered impact through efficient energy losses in any case \citep{2021ApJ...922L...3L}.

\begin{figure}
\includegraphics[width=\columnwidth]{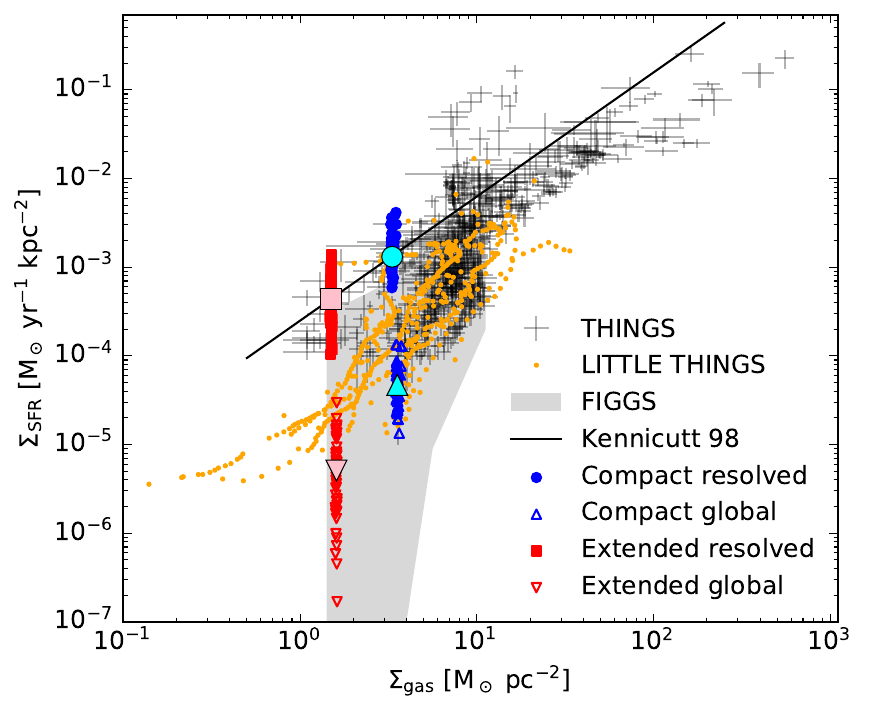}
\caption{The SFR surface density as a function of gas mass surface density in 10 Myr steps. The compact and the extended dwarf galaxies are shown in blue and red. The lower sets of open triangles show the global averages within $R=1.5$ kpc in each 10 Myr spaced snapshot while the higher sets of filled symbols have been computed as the mean over non-zero pixels in a map with 100 pc per pixel to minimize the effect of large empty regions without star formation. The pink and cyan symbols show the median value for each set of data points. The observed data are from the THINGS survey in black, \citep{2008AJ....136.2782L}, the LITTLE THINGS survey in orange \citep{2015ApJ...805..145E}, 5-95 percentile range of the dwarf galaxies in the FIGGS survey in shaded gray \citep{2015MNRAS.449.3700R}, and the line shows the standard relation of $\Sigma_\mathrm{SFR}=2.5\times 10^{-4} \Sigma_\mathrm{gas}^{1.4}$ from \citet{1998ApJ...498..541K}. \label{fig:KS}}
\end{figure}

\subsection{$\Sigma_\mathrm{SFR}$ vs. $\Sigma_\mathrm{gas}$}

To verify that the simulations produce a realistic star formation environment even when the star formation efficiency per free fall time parameter has been replaced with a strict Jeans-mass dependent threshold, we compare the SFR and gas mass surface densities ($\Sigma_\mathrm{SFR}$ and $\Sigma_\mathrm{gas}$) to observed values in Fig. \ref{fig:KS}. The blue and red open data points located at lower $\Sigma_\mathrm{SFR}$ show the two galaxy models that include stellar winds in 10 Myr steps using globally averaged values of $\Sigma_\mathrm{SFR}$ and $\Sigma_\mathrm{gas}$ over the central 1.5 kpc. The resolved values are shown with the filled circles located above the global values, estimated as an average over maps projected in the x--y plane with a resolution of 100 pc per pixel, and by excluding pixels with no data. Fig. \ref{fig:KS} also includes the median values of both the global and the resolved estimates. The underlying data show the THINGS survey that includes local spiral galaxies \citep{2008AJ....136.2782L}, the LITTLE THINGS survey that consists of local dwarf galaxies \citep{2012AJ....144..134H, 2015ApJ...805..145E}, 5-95 percentile range of the dwarf galaxies in the FIGGS survey \citep{2008MNRAS.386.1667B, 2015MNRAS.449.3700R}, and the traditional \citet{1998ApJ...498..541K} relation. 

The global values in the simulated dwarf galaxies show good agreement with the variety of observed values for dwarf irregular galaxies. The resolved values capture only the star-forming regions, as if the entire galactic disk formed stars at the localized rate. In reality, the dwarfs do not form stars across the entire disk, therefore the global average dilutes the value of $\Sigma_\mathrm{SFR}$. The median values of the resolved $\Sigma_\mathrm{SFR}$ estimates fall exactly on the Kennicutt relation, demonstrating that on small spatial scales the simulated dwarfs do form stars according to relations typically observed in more massive galaxies that form stars continuously at higher rates. 

The global results in Fig. \ref{fig:KS} are also similar to previous studies that simulated the same galaxy models at higher metallicity (0.1 \zdot) with and without the star formation efficiency per free-fall time parameter \citep{2017MNRAS.471.2151H, 2022MNRAS.509.5938H}. The only difference is that our lower metallicity dwarf galaxies form slightly less stars due to less efficient cooling. The results in Fig. \ref{fig:KS} demonstrate a self-regulated galactic star formation environment. This is enabled by the implementation of an accurate non-equilibrium cooling model, a reasonably high star formation threshold and a physically motivated time scale for the unresolved final cloud collapse ($t_\mathrm{dyn}$ and Fig. \ref{fig:tdyn}), that are combined with detailed stellar feedback prescriptions including, most importantly, pre-SN components such as photoionizing radiation and stellar winds \citep{2017MNRAS.471.2151H, 2021MNRAS.504.1039R}. 

\begin{figure}
\includegraphics[width=\columnwidth]{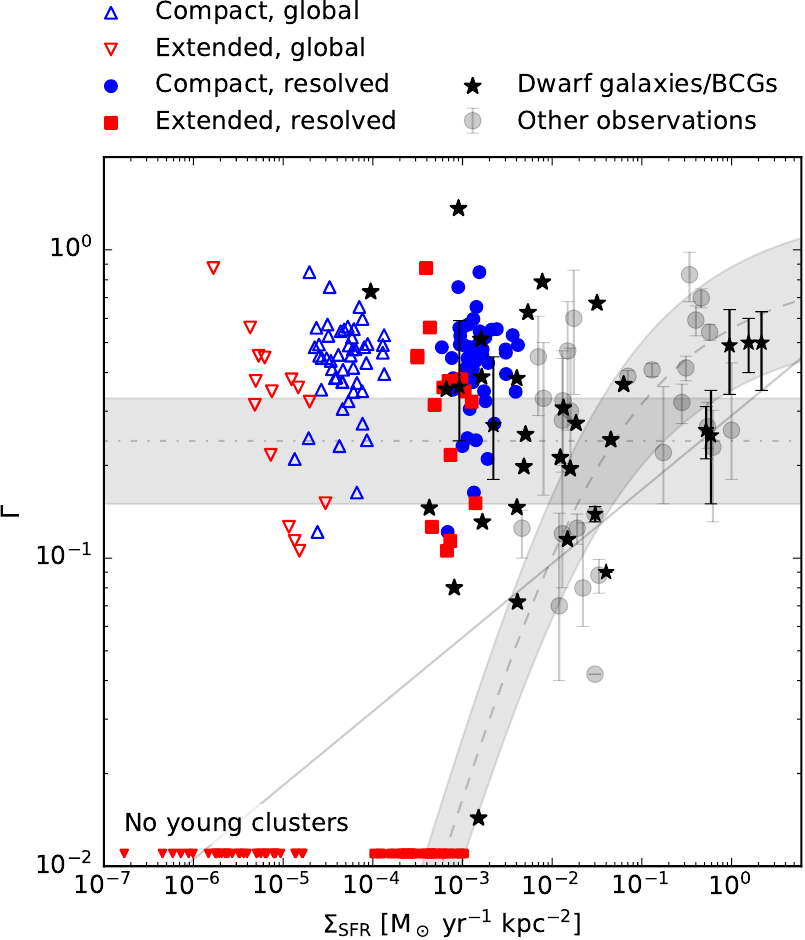}
\caption{The formation efficiency of young ($<10$ Myr) bound clusters more massive than 100 \mdot{} as a function of the SFR surface density. The compact and the extended dwarf galaxies are shown in blue and red, respectively. The $\Sigma_\mathrm{SFR}$ shows the global (open) and resolved (filled) values from Fig. \ref{fig:KS}. The star symbols (global $\Sigma_\mathrm{SFR}$) and crosses (resolved $\Sigma_\mathrm{SFR}$) at the bottom are snapshots in the extended dwarf with no young clusters. The observational reference data for young clusters ($<10$ Myr) in dwarf galaxies and blue compact galaxies (BCGs), highlighted in black, are from \citet{2011MNRAS.417.1904A}, \citet{2011AJ....141..132P}, \citet{2012ApJ...751..100C}, \citet{2017ApJ...849..128C} and \citet[][using $\Sigma_\mathrm{SFR}$ of young resolved stellar populations]{2023MNRAS.519.3749C}, while other observations shown in gray are from \citet{2010MNRAS.405..857G}, \citet{2011AJ....142..129A},  \citet{2015ApJ...804..123L}, \citet{2015MNRAS.452..246A}, \citet{2016MNRAS.460.2087H}, \citet{2017ApJ...849..128C}, \citet{2018MNRAS.477.1683M}, \citet[][age limit of $30$ Myr]{2019A&A...628A..60F} and \citet{2020MNRAS.499.3267A}. The best fit relation from \citet{2010MNRAS.405..857G} is shown with the solid line, the best fit constant value of $24\%\pm9$ from \citet{2017ApJ...849..128C} is shown with dot-dashed line, and the analytic model from \citet{2012MNRAS.426.3008K} with a $\pm0.2$ dex range is shown with a dashed line.}
\label{fig:gamma}
\end{figure}

\subsection{Cluster formation efficiency}

In Fig. \ref{fig:gamma} we show the cluster formation efficiency $\Gamma$ in the two galaxy models with winds. We define $\Gamma$ as the ratio between mass in young bound clusters with ages younger than 10 Myr, and stars formed in the past 10 Myr. Following the methods prescribed in the observational literature and the reference data in Fig. \ref{fig:gamma}, we only include clusters more massive than 100 \mdot. For completeness we also show snapshots that have no young clusters in Fig. \ref{fig:gamma}. Due to the bursty SFR, the extended dwarf galaxy does not occasionally have any bound young star clusters. We compare to a range of observed values of $\Gamma$ bearing in mind it is difficult to assess the boundness of star clusters in extragalactic environments (see e.g. the discussion in \citealt{2020SSRv..216...69A}).
Measurements of $\Gamma$ using young star clusters ($<10$ Myr) can suffer from contamination by unbound clusters and are often higher compared to estimates using older clusters (up to \mbox{100 Myr}) which only provide lower limits for $\Gamma$ due to cluster dissolution and disruption. Here we limit mainly to surveys that report $\Gamma$ for young clusters ($\lesssim 10$ Myr), to compare the clustered star formation process in observations and our simulations. In particular we want to minimize any discrepancies arising from our incomplete dynamical treatment which leads to less efficient cluster disruption (see Section \ref{section:massloss}).
Data for dwarf galaxies from \citet{2011MNRAS.417.1904A}, \citet{2011AJ....141..132P}, \citet{2012ApJ...751..100C} and \citet{2023MNRAS.519.3749C}, as well as results for the Magellanic clouds ($<10$ Myr estimate from \citealt{2017ApJ...849..128C}) have been highlighted with black. For $\Sigma_\mathrm{SFR}$ in the \citet{2023MNRAS.519.3749C} data we use the estimate based on the young ($<10$ Myr) resolved stellar populations. The absolute error for $\Gamma$ in \citet{2023MNRAS.519.3749C} can exceed 100\%, therefore we omit plotting the errorbars for clarity. 

Considering that our \textsc{subfind}-based young bound cluster sample is 100\% complete, we find that at least ten percent of stars form in bound star clusters in the compact dwarf. The extended dwarf has periods where no young clusters more massive than 100 \mdot{} are present, i.e. $\Gamma=0$. Larger scale simulations find typically values between a few percent and almost unity, but in higher $\Sigma_\mathrm{SFR}$ environments \citep{2019MNRAS.490.1714P, 2022MNRAS.514..265L, 2023MNRAS.519.1366G}. Our values of $\Gamma$ have a similar range as in previous studies with isolated and starburst dwarfs \citep{2020ApJ...891....2L, 2022MNRAS.509.5938H} and disk galaxy mergers \citep{2022MNRAS.514..265L} but no environmental dependence is found here as the simulations span a relatively small range in globally averaged ISM densities. The photometric detection discussed in Section \ref{section:clusters} would also not affect the results since the young cluster population is easily detected in this low-SFR face-on dwarf galaxy.

\begin{figure}
\includegraphics[width=\columnwidth]{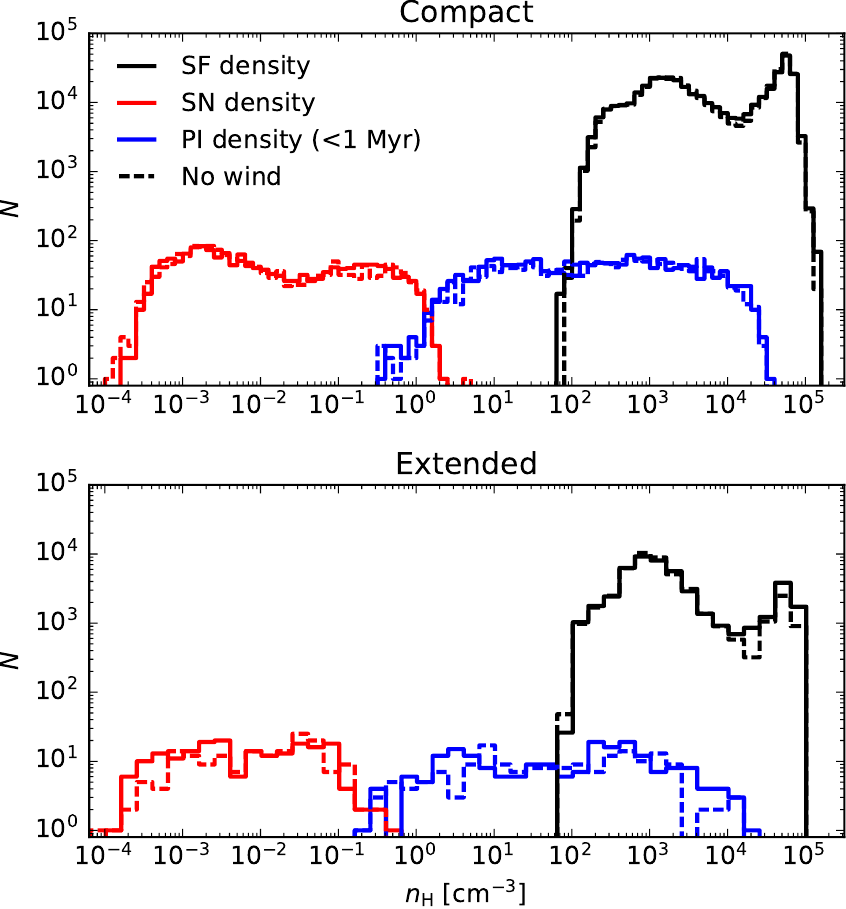}
\caption{The realized star formation threshold densities (black), the ambient densities around young photoionizing stars (blue) and the ambient densities around SNe (red). The photoionization densities are shown for stars younger than 1 Myr and the values have been stacked over all snapshots in steps of \mbox{1 Myr}. The solid and dashed lines show the runs with and without winds, and the top and bottom panels show the compact and the extended dwarf galaxy, respectively.  \label{fig:densities}}
\end{figure}

\subsection{Environmental densities of stellar feedback}

As discussed in \citet{2022MNRAS.509.5938H}, the star formation threshold affects strongly the gas densities of star formation but only weakly the densities in which SN explosions occur, as long as early stellar feedback is included. Here we exclude the star formation efficiency parameter, thus our current model is closest to the 0\% efficiency model in \citet{2022MNRAS.509.5938H}. We show the star formation densities, SN densities as well as the densities around photoionizing/wind releasing stars up to an age of 1 Myr in Fig. \ref{fig:densities}. The stars form at densities of $10^{2}$--$10^{5}$ cm$^{-3}$ as described by the Jeans-mass dependent threshold, followed by photoionization after a time delay as depicted in Fig. \ref{fig:tdyn} at slightly lower densities. SNe occur in significantly lower densities of less than $1$ cm$^{-3}$ due to the pre-SN feedback. The SNe in the compact dwarf galaxy have peak densities slightly higher than in the extended dwarf, which is consistent with the less compact configuration of the gas disk. Otherwise the distributions of star formation and feedback densities in the two models are quite similar and in agreement with previous studies \citep{2017MNRAS.471.2151H, 2022MNRAS.509.5938H}. Differences between the runs with and without winds are negligible. 

\begin{figure}
\includegraphics[width=\columnwidth]{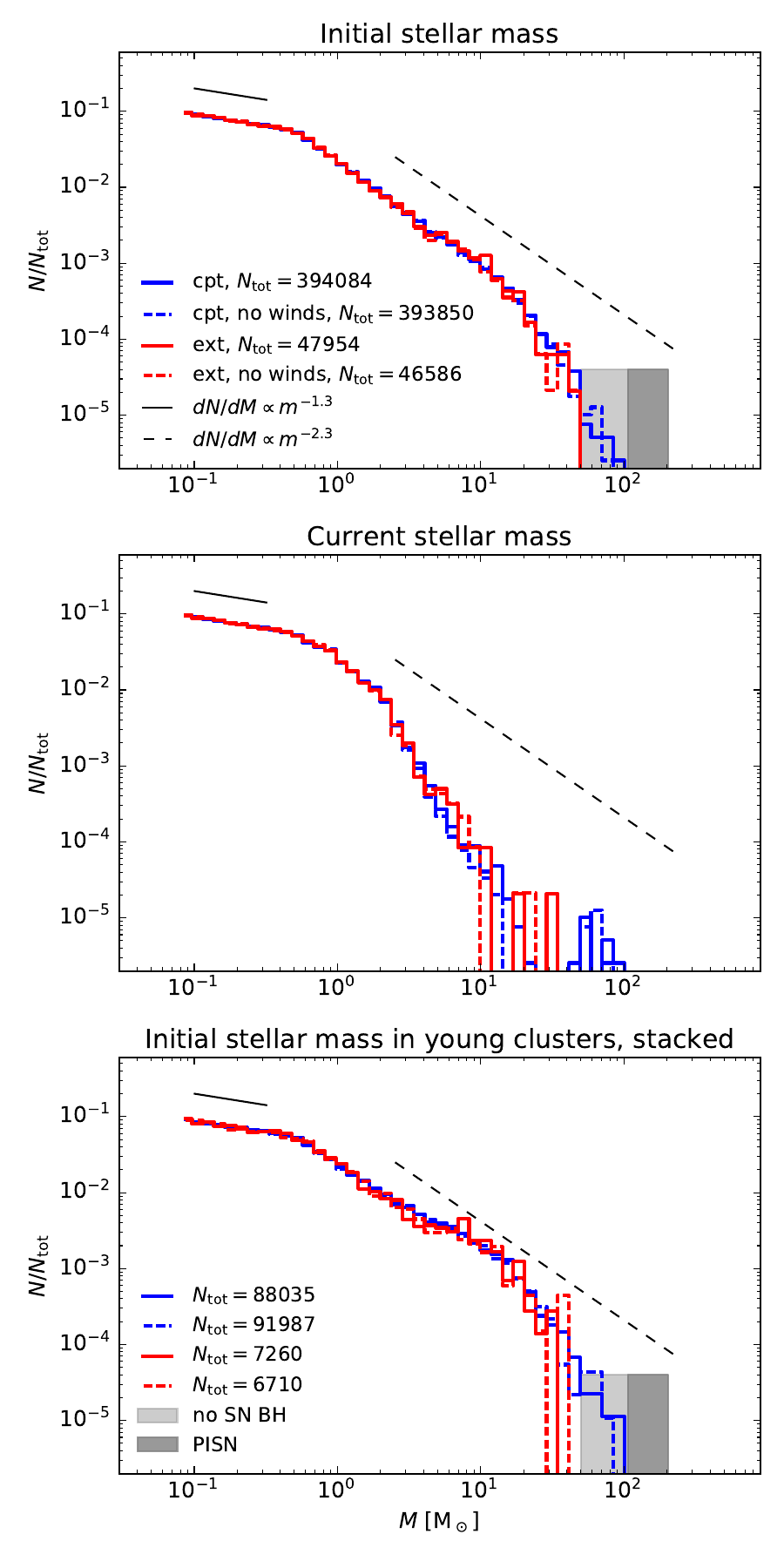}
\caption{Top: the initial masses of stars present in the final snapshot of the compact (blue) and the extended (red) dwarf galaxies. The solid and dashed coloured lines show the runs with and without stellar winds. Stars are realized down to 0.08 \mdot, and the highest stellar mass is limited by the local reservoir available when the stochastic sampling is performed. The mass functions have been normalized by the total number of stars to ease the comparison and the total number of stars is indicated in the legend. The thin solid and dashed lines show power-law mass function slopes of $-1.3$ and $-2.3$ and the shaded regions indicate the initial mass range of stars that collapse directly into black holes (light) and stars that explode as PISNe with no compact remnants (dark). Middle: the current mass of all stars in the final snapshot, taking into account the mass lost in stellar winds and SNe. Bottom: the normalized IMF of stars that are bound in young ($<10$ Myr) star clusters, stacked in 10 Myr steps across the runs. \label{fig:IMF}}
\end{figure}

\subsection{Stellar mass function}\label{section:imf}

In the isolated low-metallicity dwarf galaxies the range of SFRs from \mbox{$10^{-6}$ \mdot{} yr$^{-1}$} to $10^{-3}$ \mdot{} yr$^{-1}$ results in up to 1000 \mdot{} of new stellar mass to be sampled per Myr in the entire galaxy. Meanwhile, the dynamical time scale of the star-forming gas at densities beyond $n_\mathrm{H}\sim10^{3}$ cm$^{-3}$ is typically less than a Myr as shown in Fig. \ref{fig:tdyn}. Therefore,  the accumulated mass reservoir in the regions with a search radius of 1 pc where the IMF sampling is performed never actually allows stellar masses beyond $\sim100$--200 \mdot. In addition, there is in total $2\times 10^5$ \mdot{} and \mbox{$2.4\times 10^4$ \mdot{}} of stars in the compact and the extended dwarf galaxies after 500 Myr of star formation, respectively. The sample size of stellar masses in the extended dwarf is smaller, therefore it is also statistically less likely that stellar masses in the steep high-mass end get sampled even if they might be rejected by the sampling routine. After 500 Myr of isolated evolution, the highest initial stellar mass was 146 \mdot{} in the compact dwarf, and 40.7 \mdot{} in the extended dwarf.

The top and middle panels of Fig. \ref{fig:IMF} show the initial (top panel) and final (middle panel) masses of all stars in both of the runs, with and without stellar winds, at the end of the simulations. We also indicate the initial mass range of stars that collapse directly into black holes as well as the range of masses that, if realized, explode as PISNe and leave no compact remnant (the one star with 146 \mdot{} is therefore not shown). Stars with initial masses above $50$ \mdot{} remain as stellar mass black holes but with significantly larger masses compared to the compact objects left behind by the core collapse SNe with initial masses below $50$ \mdot{}. The more massive black holes are located in the high-mass end of the middle panel of Fig. \ref{fig:IMF}. 

In the present implementation, the simulations undersample the input Kroupa IMF at the high mass end. As a result, we have in total $\sim15$\% fewer massive ($>8$ \mdot) stars by number compared to a pure Kroupa IMF between 0.08 \mdot{} and 500 \mdot, and in total 8\% fewer SN events ($<50$ \mdot). The mass functions in the two runs are fairly similar, especially considering the low-number statistics in the high-mass end. The models with and without winds do not show differences, which is consistent with previous numerical studies (e.g. \citealt{2008MNRAS.391....2D}).

The bottom panel of Fig. \ref{fig:IMF} shows the IMF of stars that are part of bound clusters younger than 10 Myr. We stack the young clusters over the 500 Myr simulation time in steps of 10 Myr due to the low number of clusters at any given time especially in the extended dwarf galaxy. There is a slight over-abundance of massive stars, at $\sim5$--\mbox{$20$ \mdot} or so, in the young clusters compared to the overall $-2.3$ slope of the IMF. Such early mass-segregation indicates that the massive stars either form in clusters or migrate inward during the early evolution of the clusters. As we sample from a Kroupa IMF, the only way for clusters to form initially with a super-Kroupa (top-heavy) high-mass end would be for the massive stars to inhibit the realization of the rest of the IMF. At relatively low rate of star formation, if a massive star was to form and exhaust the local reservoir, the region around would only be able to proceed to form low-mass stars. Interestingly, \citet{2022MNRAS.515..167G} found a similar excess of massive stars compared to a pure Kroupa IMF in a star cluster forming in a cloud-scale simulation, without imposing any pre-defined input IMF.

\begin{figure}
\includegraphics[width=\columnwidth]{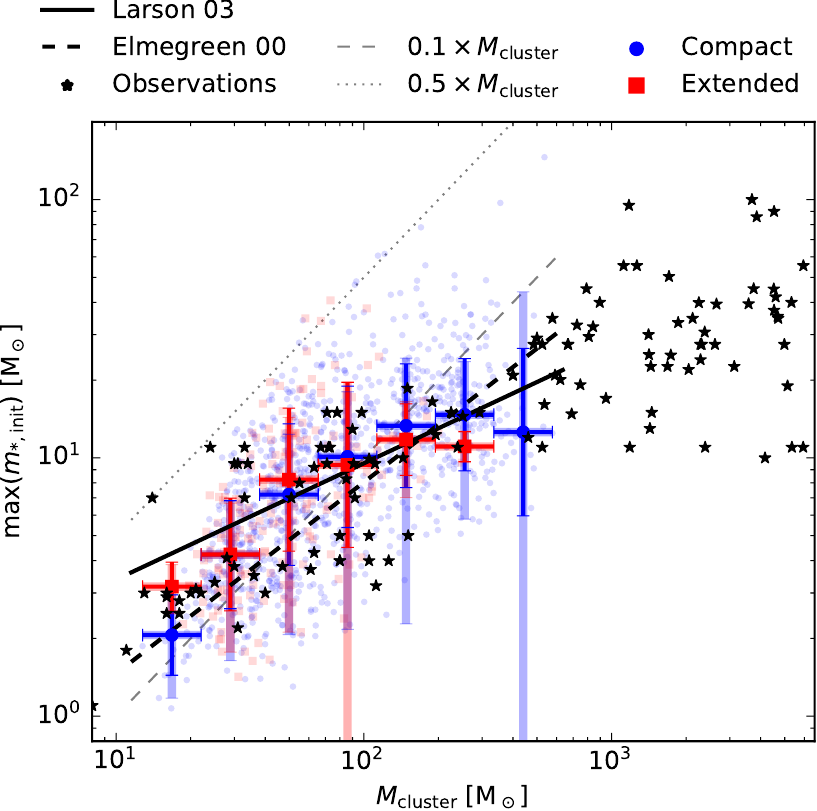}
\caption{The highest initial stellar mass in each young star cluster (\mbox{$<5$ Myr}) stacked over the two simulations (compact in blue, extended in red) in steps of 5 Myr. The coloured errorbars show the binned median and standard deviation in the maximum stellar mass, computed in linear space (light errorbars) and logarithmic space (dark errorbars). The black stars show data for single young clusters (\mbox{$<3$ Myr}) collected in \citet{2013MNRAS.434...84W}. The black solid line is an observed relation from \citet{2003ASPC..287...65L} and the dashed black line is the analytic solution from \citet{2000ApJ...539..342E}. The gray diagonal lines indicate the mass fraction of 10\% and 50\% of the total mass of each cluster comprised by the highest mass star. \label{fig:mmax}}
\end{figure}

\subsection{Highest stellar mass in clusters}

In Fig. \ref{fig:mmax} we show the relation between the highest initial stellar mass and the cluster mass in the simulated clusters of the two simulations that include stellar winds. We stack clusters younger than \mbox{5 Myr} in steps of \mbox{5 Myr}, since observed reference data are for young star clusters with estimated ages of up to a few Myr. The observed data points have been taken from Table A1 of \citet{2013MNRAS.434...84W} and the observed relation with $m_\mathrm{*,max}\propto M_\mathrm{cluster}^{0.45}$ is from \citet{2003ASPC..287...65L}. The analytic estimate from \citet{2000ApJ...539..342E} with $m_\mathrm{*,max}\propto M_\mathrm{cluster}^{1/1.35}$ is recovered by integrating the high-mass end of a Salpeter IMF \citep{1955ApJ...121..161S}. More sophisticated models in \citet{2000ApJ...539..342E} (see also e.g. \citealt{2004MNRAS.348..187W, 2009ApJ...691..823H}) introduce an upper-limit cut-off towards higher stellar masses, to account for the fact that currently largest observed stellar masses are in the 100--200 \mdot{} range (e.g. \citealt{2016MNRAS.458..624C, 2018Sci...359...69S}). 

\begin{figure}
\includegraphics[width=\columnwidth]{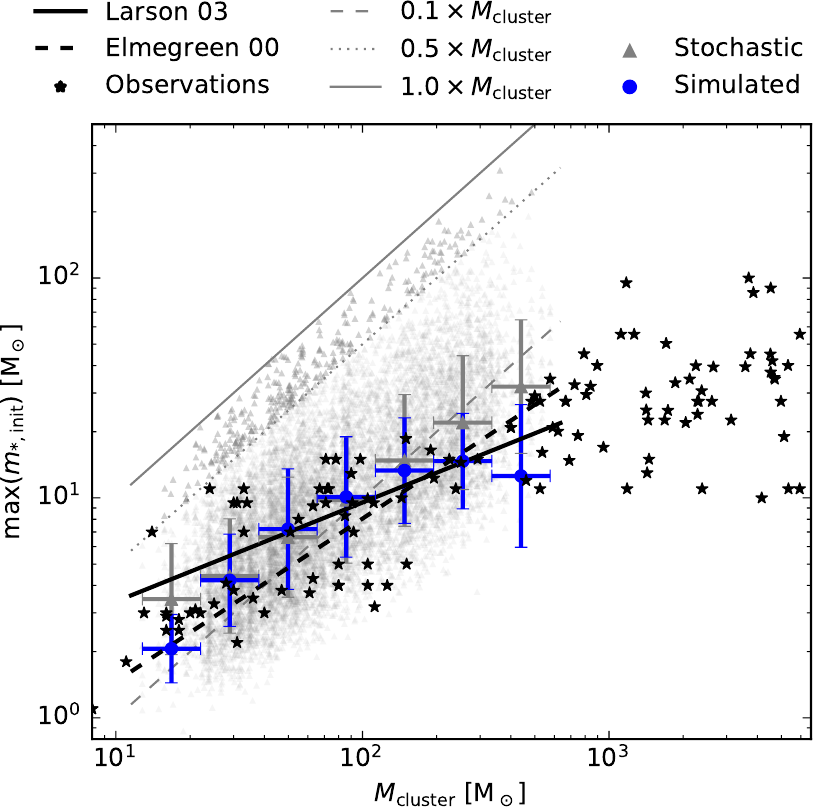}
\caption{The highest initial stellar mass as a function of the star cluster mass in young star clusters formed in the compact dwarf galaxy (blue logarithmic errorbars from Fig. \ref{fig:mmax_MC}) compared to a stochastically sampled IMF. The gray triangles in the background and the gray errorbars show the results from 10 randomly drawn IMF realizations for each cluster in the compact dwarf galaxy. The points above $\mathrm{max}(m_*)=0.5M_\mathrm{cluster}$ have been highlighted with a darker gray for visibility. The gray diagonal lines indicate the mass fraction of 10\%, 50\% and 100\% of the total cluster mass comprised by the highest mass star, the black stars and lines show the same observed data and analytic models as in Fig. \ref{fig:mmax}.} \label{fig:mmax_MC}
\end{figure}

As expected, our simulations result in an increasing trend between the highest initial stellar mass and the total mass of the cluster. The binned median values from our stacked star clusters follow  a similar relation with the observed young clusters. To quantify the contribution of pure stochasticity, we show in Fig. \ref{fig:mmax_MC} the corresponding results for randomly sampled star cluster IMFs. For each young cluster in the compact dwarf galaxy, we sample the IMF randomly 10 times using the same Kroupa IMF that is used as input in our simulations. The relation in the binned stochastic sample follows a similar trend as the \citet{2000ApJ...539..342E} analytic solution that uses a Salpeter IMF. The binned data of our simulated clusters are comparable to the stochastically populated clusters between cluster masses of a few 10 \mdot{} and 100 \mdot{}, and elsewhere the simulated clusters fall below the stochastic sample. The standard deviation of the simulated sample is also comparable or smaller than in the stochastic sample. The maximum initial stellar mass can reach the cluster mass when the IMF is populated stochastically, whereas the simulated star clusters (as shown in Fig. \ref{fig:mmax}) barely reach $\mathrm{max}(m_*)=0.5\times M_\mathrm{cluster}$ in agreement with the observed data points. The simulated clusters have to be composed of at least 50 stars, therefore one star cannot by definition comprise the entire mass of the cluster.

Our implemented locally mass-conserving stellar mass sampling routine differs from the pure stochastic sampling in low and high mass star clusters. Together with the IMF in Fig. \ref{fig:IMF}, we see that the number of massive stars is limited globally by the star formation environment of the simulated dwarf galaxies. However, comparing to observations, it is difficult to differentiate whether the stochastic sample or the simulated clusters agree better with the observed data when the cluster masses are limited to \mbox{$<1000$ \mdot}. Observations have shown indication both for \citep{2009ApJ...695..765M, 2009ApJ...706..599L, 2018MNRAS.477.5554W} and against \citep{2011ApJ...741L..26F, 2012ApJ...744...44W} a top-light IMF in galaxies with low star formation activity and a clear consensus of whether the IMF should be universal across environments remains under debate \citep{2010ARA&A..48..339B}. In our simulations, higher localized SFR densities (as realized at a higher metallicity or in a starburst) would result in more high-mass stars and a more Kroupa-like IMF, reminiscent of models that employ optimal rather than purely stochastic sampling (e.g. \citealt{2017A&A...607A.126Y}).

In the context of numerical star-by-star simulations that also analyse star clusters, \citet{2021PASJ...73.1036H} find a similar relation between the highest mass star and the host cluster mass. Based on the results in \citet{2021PASJ...73.1036H}, the spread in the maximum stellar mass -- cluster mass relation found in our simulated star cluster sample could be tightened e.g. by introducing a more stringent, density dependent searching radius in the IMF sampling. Such a density dependent reservoir would result in a stronger constraint for the maximum stellar mass in each star-forming region.

\begin{figure*}
\includegraphics[width=\textwidth]{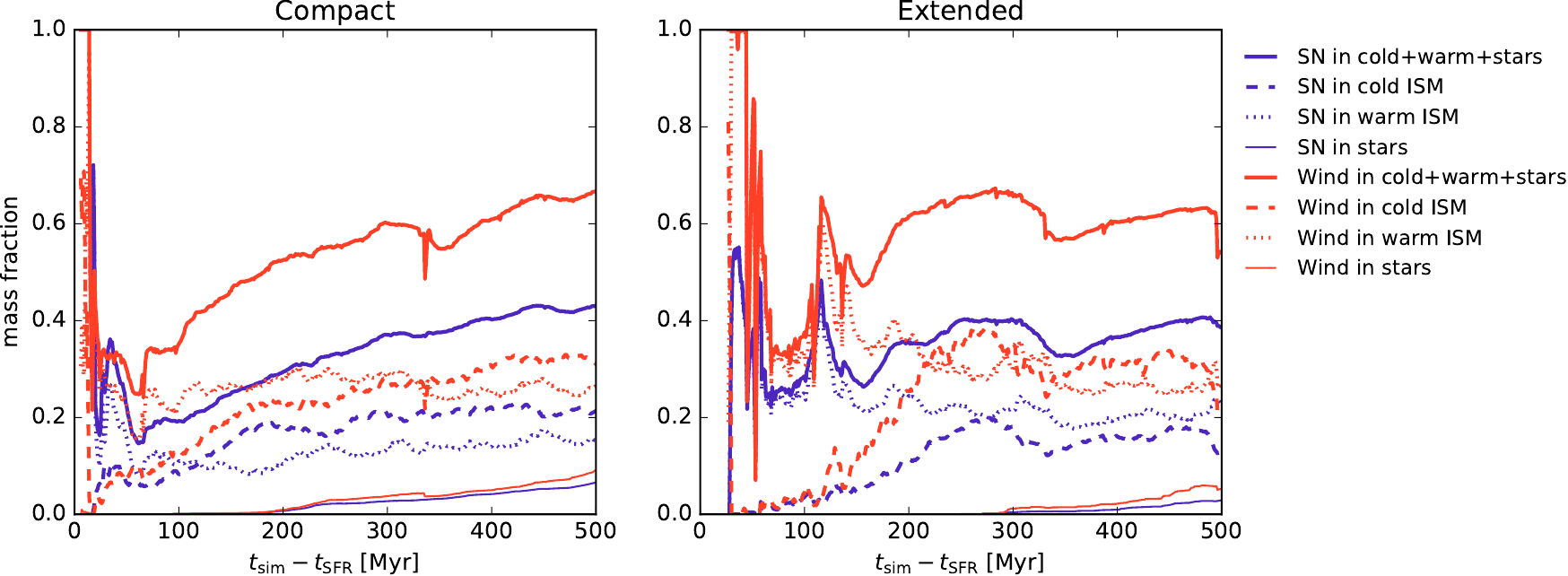}
\caption{The cumulative enrichment history of the ISM and the new stars via SNe (red) and winds (blue) in the compact (left) and the extended (right) dwarf galaxy. The mass fraction of produced SN and wind material located in the cold ISM ($<300$ K, dashed), warm ISM ($300<T<2\times 10^4$ K, dotted) and stars (thin solid) computed relative to the total mass produced by each enrichment process. The ISM is defined as gas within $\pm 300$ pc of the galactic midplane. The thick solid line shows the sum over the cold and warm ISM and the stars, and the remainder of the material is located either in the hot gaseous phase or outside of the galactic disk. \label{fig:ISM}}
\end{figure*}

\section{Metal enrichment}\label{section:enrichment}

\subsection{Formation of enriched stars}

In the following we analyse the fate of the wind and SN material that end up captured in the interstellar gas, thus leading to the formation of enriched stars. Enriched gas is defined as particles that have received any wind and/or SN material, and enriched stars are stars formed out of enriched gas regardless of their later evolution. We individually track each 13 elements released in winds and SNe separately, which allows us to study which enrichment process contributed to the initial elemental abundance of each individual star.

Fig. \ref{fig:ISM} shows the cumulative enrichment history via winds and SNe for the ISM and the newly formed stars. We show the cumulative fraction of wind and SN mass located in the cold ($<300$ K) and the warm ($300$ K $<T<2\times 10^4$ K) phases of the ISM. The ISM is defined as all gas particles located between $\pm300$ pc of the midplane. The mass fraction of wind and SN material locked in new stars is indicated as well. The extended dwarf is more prone to stochastic fluctuations in the fraction of feedback material located in different phases of the ISM. Bursts of feedback can drive more efficient outflows (as shown later) in the extended disk that is less enriched to begin with due to the lower total SFR. The sum of the material in stars, in the cold ISM and in the warm ISM shows how up to $\sim40$\% of the SN and $\sim65$\% of the wind material are located either in stars or available to star formation in the cold and warm ISM. The rest, i.e. $\sim60$\% of the SN and $\sim35$\% of the wind material are either in the hot gas phase or in galactic outflows and therefore not available for enriched star formation. Approximately 20\% of the SN and 30\% of the wind material are locked in the cold ISM and up to 6\% of the SN and 9\% of the wind material end up locked in new stars by the end of the simulations. Winds are therefore slightly easier to capture in the star-forming gaseous phase, compared to the initially very hot SN material.

\begin{figure*}
\includegraphics[width=\textwidth]{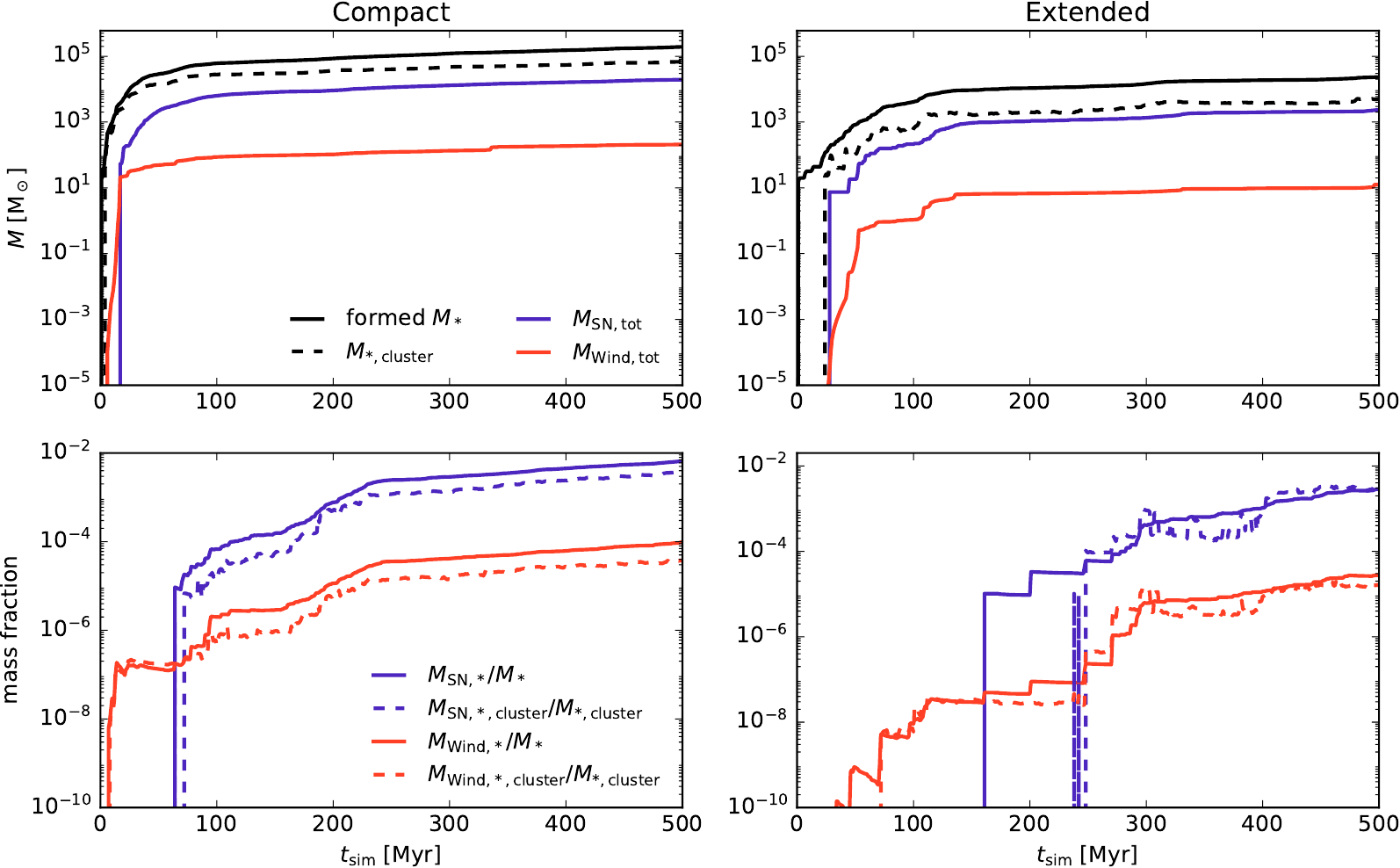}
\caption{Star formation and enrichment histories of the compact (left) and extended (right) dwarf galaxies with stellar winds. Top: the cumulative stellar mass ($M_*$, black solid), mass in bound star clusters ($M_\mathrm{*,cluster}$, black dashed), cumulative mass ejected by SNe ($M_\mathrm{SN,tot}$, blue) and  winds ($M_\mathrm{wind,tot}$, red). Bottom: the fraction of stellar mass comprised of wind and SN material in all stars (solid) and in stars bound to clusters (marked by the underscript $_\mathrm{cluster}$, dashed lines).\label{fig:ejecta}}
\end{figure*}

The cumulative mass of new stars ($M_*$), total mass of bound star clusters (>50 stars, $M_\mathrm{*,cluster}$), total produced wind mass (\mbox{$M_\mathrm{wind,tot}=M_\mathrm{wind}$} for easier separation from the other quantities) and the total produced SN mass ($M_\mathrm{SN,tot}=M_\mathrm{SN}$ likewise) are shown in the top row of Fig. \ref{fig:ejecta}. The cluster mass is computed for each snapshot separately. Once massive stars form, winds are in general the first enrichment process to kick in, followed by SNe with a delay. On the right (the extended disk), the first massive star had a mass just below our wind table limit of 9 \mdot, therefore the first SN occurs without wind enrichment, followed by more massive star formation which then start releasing winds. In the left hand panel, the jump in wind mass and respective drop in the fraction of winds locked in stars at $t_\mathrm{sim}\sim335$ Myr corresponds to the short lifetime of the \mbox{146 \mdot{}} star. The explosion of the PISN is also reflected in a brief drop in wind fraction located in the cold phase (Fig. \ref{fig:ISM}), as the PISN heats and blows out the surrounding medium which the progenitor stars previously enriched via winds. 

The bottom row of Fig. \ref{fig:ejecta} shows the total mass fraction in stars comprised by the wind and SN material that are locked in the stars. We show both the overall mass of the wind and SN mass locked in stars relative to the total stellar mass, as well as the wind and SN material locked in cluster stars relative to the total star cluster mass. The mass fraction in clusters is computed separately in each snapshot and accounts for cluster mass-loss, therefore we relate the wind and SN mass in clusters to the current total cluster mass instead of the total stellar mass. The cumulative mass fraction of wind material increases gradually and reaches $10^{-5}$--$10^{-4}$ in the simulations globally. Except in the earliest phases, a slightly lower fraction of the stellar mass is comprised by wind and SN material in star clusters, compared to the global average. \citet{2019ApJ...871...20S} studied the wind retention within young, low-metallicity massive star clusters, combining similar stellar models to ours with very massive stars and one-dimensional hydrodynamical semi-analytic calculations. They concluded that clusters would need to be initially at least $10^5$ \mdot{} to retain more than a fraction of a percent of the ejected wind mass. They did not model the further star formation explicitly and assumed that the feedback releasing stellar population is single age and spherically symmetric. \citet{2021ApJ...922L...3L} showed in $10^5$ \mdot{} molecular cloud simulations how the stars in the resulting cluster can contain a mass fraction of at least $10^{-4}$ in stellar winds. \citet{2021ApJ...922L...3L} did not include radiation or SN feedback and their final star cluster masses were a factor of ten higher compared to the most massive clusters in our present study. Our results, with a self-consistently evolving population of star clusters up to $10^3$ \mdot, are therefore in broad agreement with previous more constrained studies.

In our simulations, once stars enriched by SN material start forming, they immediately exceed the mass fraction comprised by stellar wind material. However, despite the (globally) relatively brief delay between the onset of winds and SNe, the SN material appears in the enriched stars significantly later compared to the wind material. The delay between the formation of the wind enriched stars and the formation of the SN enriched stars is more than 50 Myr in the compact dwarf and 90 Myr in the extended dwarf. In general, comparing the compact and the extended dwarf galaxies, a sightly smaller fraction of all of the feedback material gets locked in stars in the extended dwarf run, as shown in Fig. \ref{fig:ISM} as well.

\begin{figure*}
\includegraphics[width=0.9\textwidth]{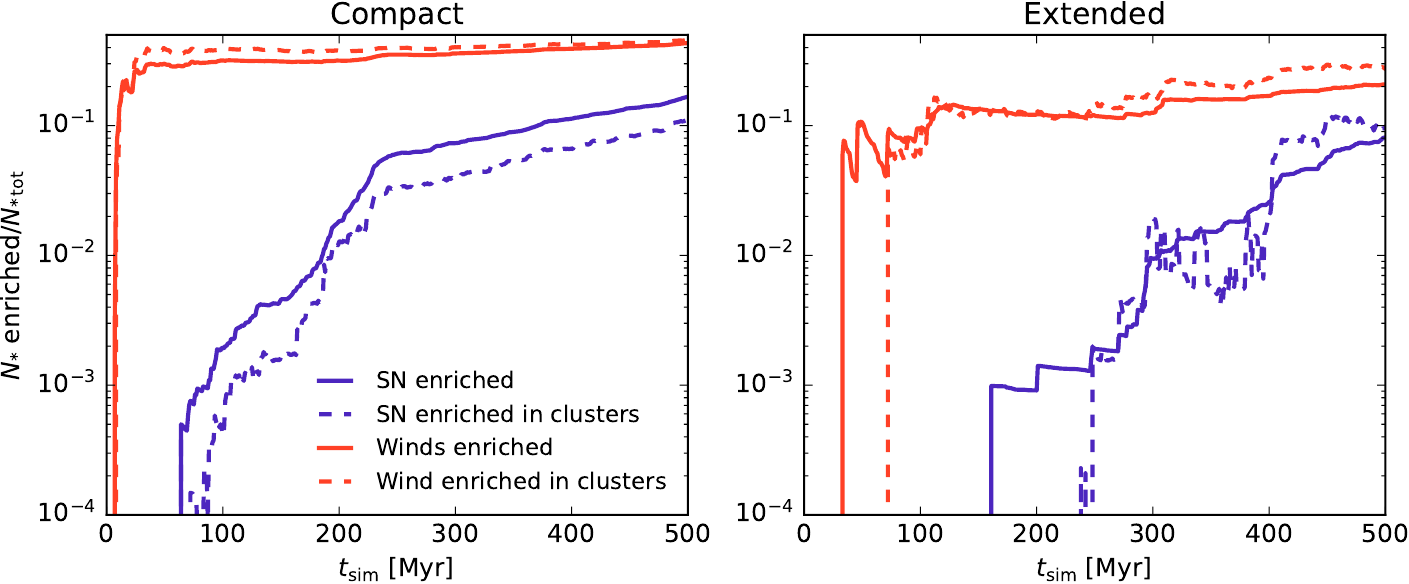}
\caption{The fraction of stars by number that formed out of gas enriched by winds (red) and SNe (blue) in the compact (left) and the extended (right) dwarf galaxy. The solid lines show the fraction of all enriched stars, while the dashed lines show the fraction of enriched stars in bound clusters related to the total number of cluster stars. \label{fig:ejectaN}}
\end{figure*}

The mass fraction of the SN mass locked in stars is almost two orders of magnitude higher than the wind mass fraction, reflecting the fact that up to $10\%$ of the stellar mass is returned as SN ejecta compared to $0.1\%$ in winds. To reach higher mass fractions of wind enrichment in new stars, more massive stars in number and with higher stellar masses are needed. Compact star cluster-forming regions, such as those presented in \citet{2020ApJ...891....2L}, might be able to form more massive stars and retain more of the early released stellar wind material. We will investigate the cluster-by-cluster enrichment of the different enrichment channels in future work. Importantly, in the bottom panel of Fig.\ref{fig:ejecta}, the delay of SN enrichment is further prolonged when stars in clusters are considered. In the extended dwarf galaxy, another 50 Myr passes before the first SN enriched star clusters form. Meanwhile, wind enriched star clusters appear almost immediately after star formation begins in the compact dwarf, and few tens of Myr delayed in the extended dwarf. The delay of SN enriched star formation and, especially, cluster formation, compared to wind enriched star formation may be a pathway to forming pristine star clusters that are only enriched by early stellar feedback.

\subsection{Enriched stars by number}

The wind and SN ejection methods differ fundamentally in how much gas and new stars by number they are able to enrich. Winds are released continuously into the $(8\pm 2)\times12$ \healpix-selected neighbours during the stellar lifetime. The neighbours are not always the same on each time step, therefore hundreds of particles get enriched by winds as opposed to the single injection into $(8\pm 2)\times12$ neighbours in each SN event. In Fig. \ref{fig:ejectaN} we show the fraction of all stars and stars in bound clusters that formed out of gas enriched by winds or SN material. Fig. \ref{fig:ejectaN} shows that the number of stars that were enriched by winds is a few times higher than that of stars enriched by SNe. The same is true for stars in clusters. Once wind enriched stars start forming, the fraction of enriched stars rapidly increases to 30--40\% in the compact dwarf and 20--30\% in the extended dwarf. The fraction of wind enriched stars evolves only gradually with time, while the SN enriched stars build up over time to comprise $\sim$8--16\% of the stars by the end of the two simulations. As SN events occur immediately after a stellar track ends, most of the SN enriched stars have also been enriched by winds.

The finding that more stars by number are enriched by winds compared to SNe has important implications for the interpretation of how enrichment occurs in star cluster-forming regions. In observed star clusters, the amount of enrichment by SN ejecta is often low or non-existent. Meanwhile, especially in massive globular clusters, a large fraction or even the majority of the stars are observed to have been enriched only with light elements and no SN ejecta. These observed results are conceptually in line with the results shown in Figs. \ref{fig:ejecta} and \ref{fig:ejectaN}: winds are locked in stars earlier than the SN material, a larger fraction of the released winds get locked in stars compared to the mass fraction of SN mass locked in stars, and the continuous nature of the winds causes the wind material to be more evenly spread across a larger number of stars.

\subsection{Wind retention in clusters}

\begin{figure*}
\includegraphics[width=\textwidth]{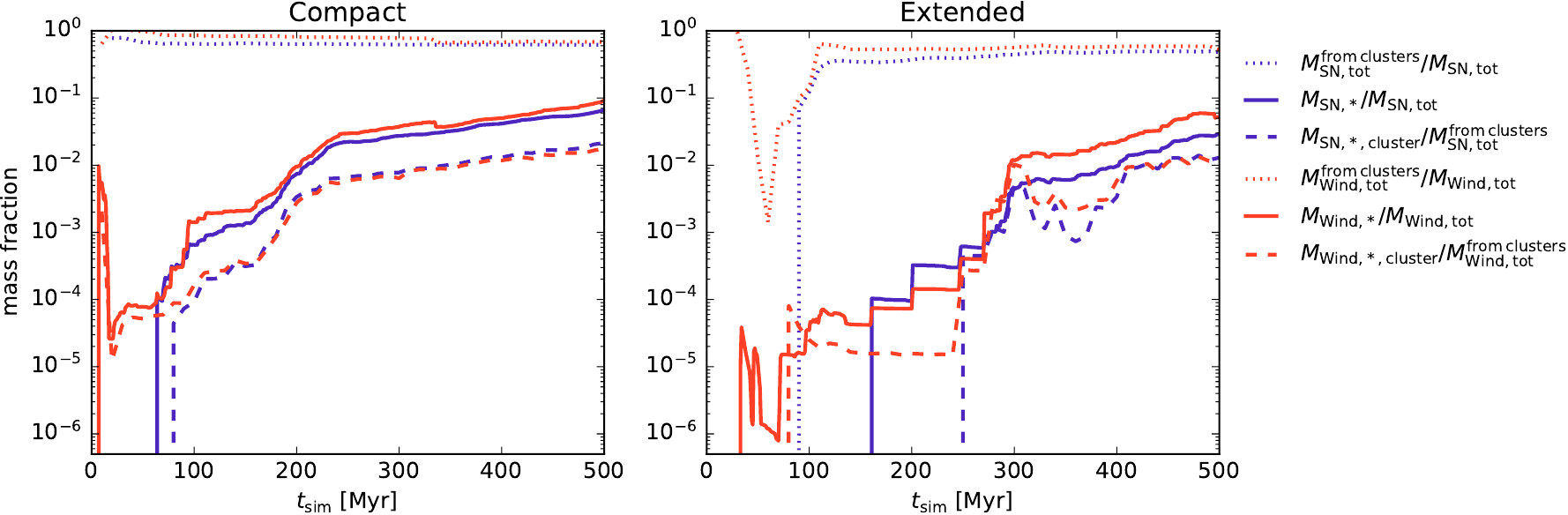}
\caption{The fraction of SN and wind mass locked in stars ($M_\mathrm{SN,*}$, $M_\mathrm{wind,*}$) and cluster stars ($M_\mathrm{SN,*,cluster}$, $M_\mathrm{wind,*,cluster}$). The solid lines show the overall mass fraction of winds and SN material locked in stars repeated from Fig. \ref{fig:ISM}, the dashed lines show the wind and SN material locked in currently bound cluster stars related to the respective cumulative mass ejected by stars in bound clusters. The dotted lines indicate the cumulative fraction of wind and SN material produced by cluster stars ($M_\mathrm{SN,tot}^\mathrm{from\, clusters}$ and $M_\mathrm{wind,tot}^\mathrm{from\, clusters}$) out of all respective  produced material. \label{fig:cl_retention}}
\end{figure*}

In Fig. \ref{fig:cl_retention} we repeat the fraction of wind material locked in stars from Fig. \ref{fig:ejecta}, and compare to the respective mass locked in stars in bound clusters relative to the mass released by massive stars in clusters. We also show the fraction of wind and SN mass released by stars that are found in bound clusters. The mass released by cluster stars is integrated in \mbox{1 Myr} steps by checking which massive stars are bound to clusters in each snapshot. In both of the runs, approximately 1--2\% of SN and wind material produced by the cluster stars ends up locked in new stars that are bound to clusters, an amount that is a few times lower than the global average.

Fig. \ref{fig:cl_retention} also shows the overall fraction of all wind and SN material released by the cluster stars. The majority of the stellar feedback is released by stars that are bound to clusters, indicating that most of the massive stars spend the majority of their lifetimes in bound clusters. This is consistent with results in Fig. \ref{fig:IMF} and Fig. \ref{fig:mmax} where we already showed how the IMF in clusters is slightly overabundant in massive stars compared to the global IMF, and that the highest mass stars are located, on average, in increasingly massive star clusters. Observations of young star clusters in the globular cluster mass range at higher redshift have also indicated that a significant fraction of the UV-radiation originates from young massive star clusters \citep{2020MNRAS.491.1093V, 2022A&A...659A...2V}, consistent with a picture where massive stars are predominantly located in star clusters.

\begin{figure*}
\includegraphics[width=\textwidth]{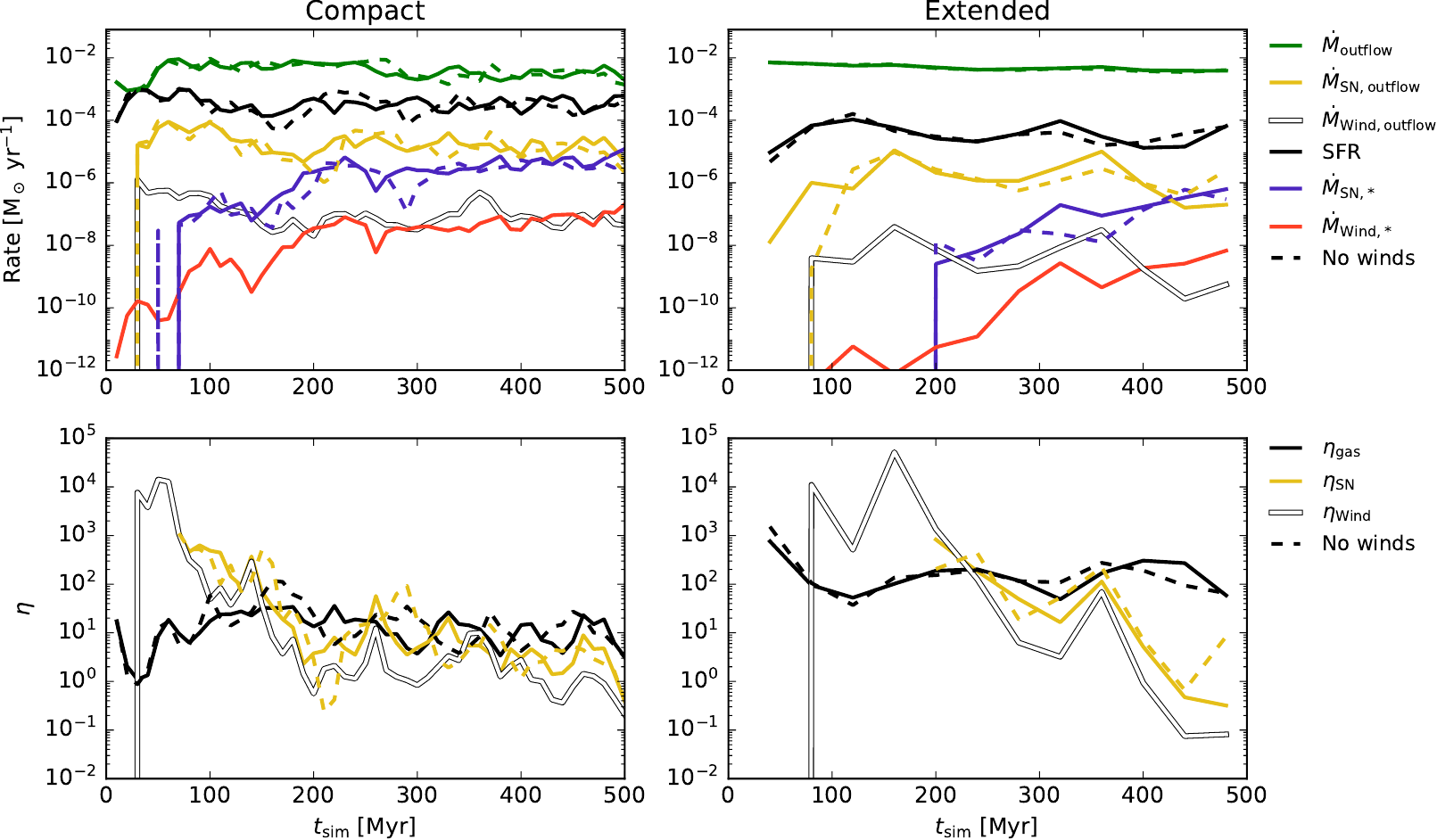}
\caption{Top: the outflow rate of gas ($\dot{M}_\mathrm{outflow}$, green), SN material ($\dot{M}_\mathrm{SN,outflow}$, yellow) and wind material ($\dot{M}_\mathrm{wind,outflow}$, white) compared to the SFR (black), the rate at which SN ($\dot{M}_\mathrm{SN,*}$, blue) and wind material ($\dot{M}_\mathrm{wind,*}$, red) are locked into new stars in the isolated dwarf galaxy runs (compact on the left, extended on the right). The solid and dashed lines show the runs with and without stellar winds. The rates have been averaged over past 10 Myr in steps of \mbox{10 Myr} on the left, and over 40 Myr on the right to account for the burstier nature of star formation in the extended dwarf galaxy. Bottom: the mass loading factors computed as outflow rate per locking rate as defined in Eqs. \ref{eq:loading_gas} and \ref{eq:loading}.\label{fig:ofr}}
\end{figure*}

\subsection{Outflow rates and loading factors}

\subsubsection{Mass loading}

The stellar feedback produces gaseous galactic outflows that are 10 to 100 times larger in mass compared to the respective SFR. The gaseous outflows are comprised of predominantly warm gas with \mbox{$T=8 000$--$15 000$ K}, and enriched by stellar feedback material. Overall, the metallicity in the galactic outflows of the compact dwarf galaxy is enhanced by up to a factor of 2--3 compared to the initial metallicity of $0.016$ \zdot, while the enhancement is only a few percent in the extended dwarf galaxy.

We compute the outflow rates at height $z=1$ kpc for quantities $q$ (gas mass, SN mass, wind mass, all metals, SN metals and wind metals separately) as the sum over all particles $i$ with mass $m_{q,i}$ moving out of the midplane with velocity $v_{z,i}$ as
\begin{equation}
   \dot{M}_{q\mathrm{,outflow}} = \sum_{i=1}^{N} \frac{m_{q,i}|v_{z,i}|}{dz}
\end{equation}
in parallel slabs of thickness $dz$ above and below the disk. Following \citet{2019MNRAS.483.3363H} and \citet{2023MNRAS.521.2196A}, we use $dz=0.1\times|z|=100\,\mathrm{ pc}$ as the thickness of the slabs. In the top row of Fig. \ref{fig:ofr} we show the outflow rates of gas, wind material and SN material in \mbox{10 Myr} steps for the compact dwarf and in 40 Myr steps for the extended dwarf. The longer averaging interval for the extended dwarf is used to average out the large variations in the rate estimates caused by the bursty episodes of star formation. In all the panels, we also show the runs without winds as dashed lines. Interestingly, the outflow rates of the two models are quite similar while the other quantities are lower in the extended dwarf. The lower gas surface density (see Fig. \ref{fig:densities}) in the extended dwarf may compensate for the lower SFR and therefore lower number of SNe in allowing the gas to escape relatively easier than in the compact dwarf.

The loading factor of the gaseous outflow with outflow rate $\dot{M}_\mathrm{outflow}$ is defined as 
\begin{equation}\label{eq:loading_gas}
    \eta_\mathrm{gas}=\frac{\dot{M}_{\mathrm{outflow}}}{\mathrm{SFR}}.
\end{equation}
For wind and SN loading, and more specifically for metal loading, there exist various definitions in the literature. The loading can be computed either in relation to the production rate, or in terms of metals locked in stars. We first quantify the mass loading according to the wind and SN outflow rates ($\dot{M}_{\mathrm{wind,outflow}}$ and $\dot{M}_{\mathrm{SN,outflow}}$) in terms of the rate at which the material is locked in stars ($\dot{M}_\mathrm{wind,*}$ and $\dot{M}_\mathrm{SN,*}$), in order to use the same definition as with the gas mass loading in Eq. \eqref{eq:loading_gas}. For the wind and SN mass loading we then have
\begin{equation}\label{eq:loading}
    \eta_\mathrm{wind}=\frac{\dot{M}_{\mathrm{wind,outflow}}}{\dot{M}_\mathrm{wind,*}},\,\,\,\,\,
    \eta_\mathrm{SN}=\frac{\dot{M}_{\mathrm{SN,outflow}}}{\dot{M}_\mathrm{SN,*}} .
\end{equation}
For completeness, we also show the locking rates in the top panels of Fig. \ref{fig:ofr}. While the locking rate of gas, i.e. the SFR, remains approximately constant, in case of the winds and SN the locking rate has a cumulative nature and starts from zero when no SN or wind material is located in the cold ISM. This was shown in Fig. \ref{fig:ISM}. As a result, the wind and SN mass loading factors according to Eq. \eqref{eq:loading} decrease with time, as ISM gets more enriched with SN and wind material. We collect the gas, wind and SN mass loading factors in the bottom row of Fig. \ref{fig:ofr} at the height of 1 kpc. 

The gas outflow rates of the two models are quite similar while the SFRs differ by an order of magnitude. As a result, the mass loading of the gaseous outflow is approximately $10$ in the compact dwarf, and 100 in the extended dwarf, throughout the simulations. The wind and SN loading factors, on the other hand, evolve with time as the locking rates evolve along ISM enrichment (Fig. \ref{fig:ISM}). The wind and SN loading factors start high, as barely any of the feedback material is locked in stars as discussed in Fig. \ref{fig:ISM} and Fig. \ref{fig:ejecta}. Once the cold ISM starts to saturate with material from stellar feedback, the loading factors drop to values of 1--10 in the compact dwarf. The extended dwarf has a slightly higher fraction of the feedback material locked in the cold ISM which leads to even lower wind and SN loading factors as time progresses. Overall, there is no significant difference between the outflow rates or the loading factors in the runs with and without winds.

\subsubsection{Metal loading}

\begin{figure*}
\includegraphics[width=\textwidth]{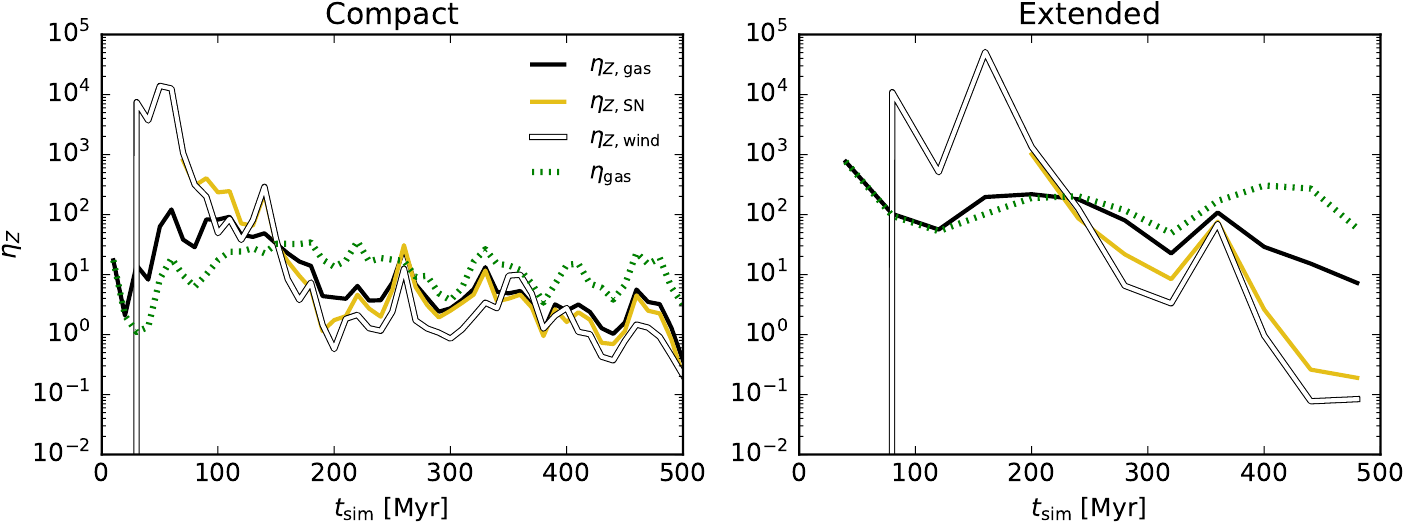}
\caption{The metal loading of all gas (black), SN metals (yellow) and wind metals (white) using the definitions in Eqs. \ref{eq:metal_loading_gas} and \ref{eq:metal_loading}. The green dashed lines shows the mass loading of gas from Fig. \ref{fig:ofr}. \label{fig:metals}}
\end{figure*}

Using the same formalism for metals (all elements heavier than He), the total metal loading factor in terms of metal outflow rate ($\dot{M}_{Z,\mathrm{outflow}}$) is given by
\begin{equation}\label{eq:metal_loading_gas}
    \eta_\mathrm{Z}=\frac{\dot{M}_{Z,\mathrm{outflow}}}{Z_\mathrm{gas}\times\mathrm{SFR}}
\end{equation}
where the denominator is the locking rate of metals into stars given by the metallicity of the star-forming gas $Z_\mathrm{gas}$ and the SFR. The wind and SN metal loading factors using the wind and SN metal outflow rates ($\dot{M}_{\mathrm{Z,wind,outflow}}$ and $\dot{M}_{\mathrm{Z,SN,outflow}}$) and locking rates ($\dot{M}_{\mathrm{Z,wind,*}}$ and $\dot{M}_{\mathrm{Z,SN,*}}$) are 
\begin{equation}\label{eq:metal_loading}
    \eta_\mathrm{Z,wind}=\frac{\dot{M}_{\mathrm{Z,wind,outflow}}}{\dot{M}_\mathrm{Z,wind,*}},\;\;\;\;\;
    \eta_\mathrm{Z,SN}=\frac{\dot{M}_{\mathrm{Z,SN,outflow}}}{\dot{M}_\mathrm{Z,SN,*}}.
\end{equation}
Similar to the wind and SN mass loading, the denominator in the metal loading is a cumulative quantity that follows the general ISM enrichment as shown in Fig. \ref{fig:ISM}. We show the metal loading factors of the simulations with winds in Fig. \ref{fig:ofr}, at the height of 1 kpc. We also indicate the total mass loading to enable easier comparison between the total metal loading and mass loading factors.

In the compact dwarf galaxy, the total metal loading factor follows the SN loading factor for the most of the time. As in the case of mass loading, at early phases the extremely low wind and SN material locking rates into new stars decouple the feedback metal loading from the total metal loading. The gas mass loading and gas metal loading are of the same magnitude, and both are clearly above unity for the most of the simulation time. Same is true for the wind and SN metal loading factors. A significantly larger amount of metals are therefore flown out of the galaxy compared to what is locked into stars in both of the simulations. This is in line with the global averages shown in Fig. \ref{fig:ejecta} and Fig. \ref{fig:ISM}, where more than 90\% of all of the wind and SN material was shown to reside outside of new stars by the end of the simulation. The metal loading factor of SN material in the outflows is also for the majority of the simulation larger than that of the wind material, enforcing the result in Fig. \ref{fig:ISM} that a clearly lower fraction of the SN material is available for enriched star formation compared to the wind material.

\begin{figure*}
\includegraphics[width=\textwidth]{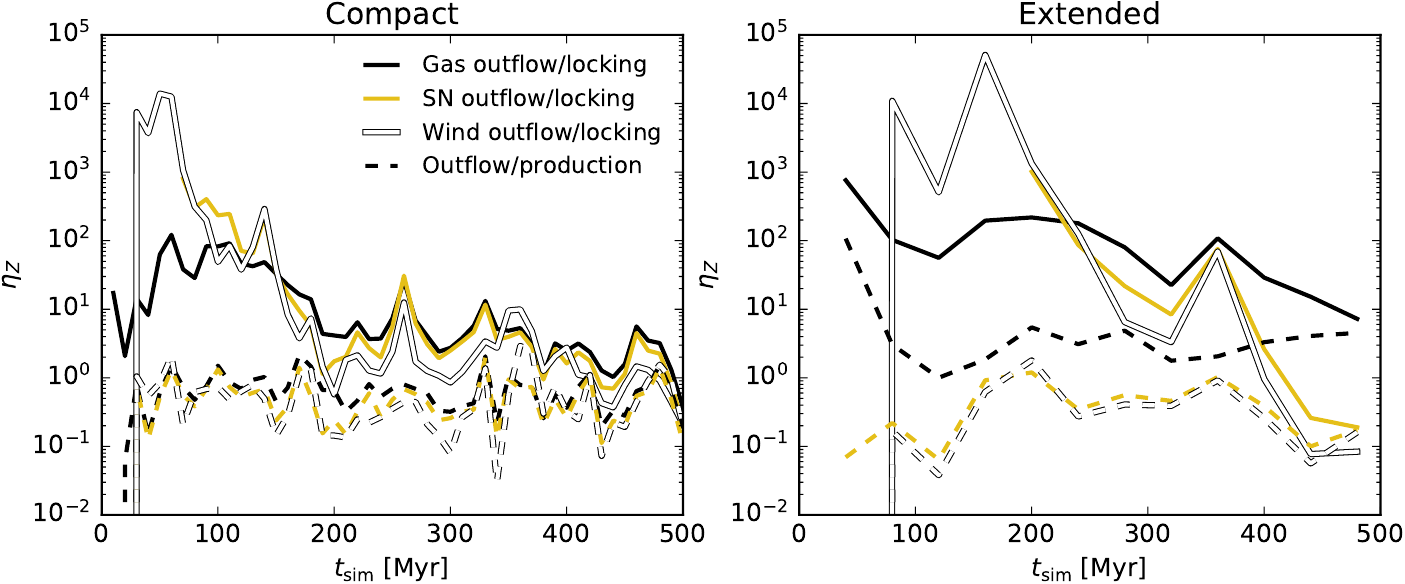}
\caption{A comparison of the two definitions for metal loading used in the literature: the solid lines show the loading factors from Eq. \eqref{eq:metal_loading_gas} and Eq. \eqref{eq:metal_loading} as well as Fig. \ref{fig:ofr}, i.e. outflow rates relative to the locking rates, while the dashed lines show the outflow rates relative to the production rates for all metals (black), for SN metals (yellow) and for wind metals (white). \label{fig:lock_vs_prod}}
\end{figure*}

To bridge the gap between different definitions of the metal loading factors, we show in Fig. \ref{fig:lock_vs_prod} the other definition for the metal loading where the \textit{locking} rates ($Z_\mathrm{gas}\times$SFR, $\dot{M}_\mathrm{Z,SN,*}$ and $\dot{M}_\mathrm{Z,Wind,*}$) in Eq. \eqref{eq:metal_loading_gas} and Eq. \eqref{eq:metal_loading} have been replaced by \textit{production} rates ($\dot{M}_\mathrm{Z,tot}$, $\dot{M}_\mathrm{Z,SN,tot}$ and $\dot{M}_\mathrm{Z,Wind,tot}$).
At low metallicity, as argued earlier, the metals in the outflows can be strongly enhanced by material released via stellar feedback. The separate SN and wind production loading factors are slightly lower than the total metal outflow rate per production, however not by more than a factor of a few. As a simple estimate, one canonical SN event releases 2.5 \mdot{} of metals and therefore increases the metallicity in a pristine ($Z=0.016$ \zdot) ambient ISM around the star by a factor of 30.  A detailed inspection of the fraction of metals in the outflow comprised of SN metals reveals that between 10\% and 60\% of the outflow metals are composed of SN material in the compact dwarf. Meanwhile, the wind mass alone is always negligible compared to the pristine metallicity and the SN metals. The total metallicity in the outflows of the compact dwarf is enhanced by up to a factor of 2--3 compared to the pristine gas. The initial injected SN material is therefore diluted by a factor of ten to result in an enhancement of only by a factor of a few in the outflowing gas metallicity. In the extended dwarf, the metal outflow per production is always above unity while the metallicity in the outflow is enhanced only by a few percent, therefore the dilution is two orders of magnitude larger.

\subsubsection{Comparison to other studies}

Here we briefly compare our main findings on the mass loading to other dwarf galaxy simulations.
As a note of caution, it is sometimes common to use an average instead of an evolving value for the denominator of the loading factors, wherein changes in the locking or production rates are not reflected in the loading factor. \citet{2022MNRAS.509.5938H} studied the mass loading factor in the same dwarf model as our compact dwarf but at higher metallicity and at a lower height of 500 pc above the disk. They arrived at mass loading factors between 10 and 100. Computing the values in our compact model at 500 pc, we arrive at values for the gas mass loading of the order of 100. \citet{2017MNRAS.471.2151H} and \citet{2021MNRAS.506.3882S} find mass loading factors at the height of \mbox{1--2 kpc} in the range of 1--10 in models that have a full feedback prescription including photoionization. \citet{2019MNRAS.482.1304E}, \citet{2023MNRAS.521.2196A} and \citet{2022arXiv220509774S} report values between 10--100 at a height of 1--3 kpc. Unlike other studies mentioned here, \citet{2023MNRAS.521.2196A} include stellar winds but exclude photoionization that in \citet{2021MNRAS.506.3882S} was found to have a significant impact in reducing the mass loading factor. 

For metal loading, \citet{2023MNRAS.521.2196A} utilized the same definition as our Eq. \eqref{eq:loading}. Their metal loading factors were of the order of 10--100, same as their mass loading, at a height of 1 kpc while our metal loading is slightly lower than our mass loading. \citet{2019MNRAS.482.1304E}, on the other hand, find even lower values of metal loading than our simulations, however at larger heights. \citet{2022arXiv220509774S} use the metal-loading factor that relates outflows to metal production, and find values of the order of 100--200, i.e. their outflows were predominantly composed of ambient metals. In Fig. \ref{fig:lock_vs_prod} we showed that for the compact dwarf the metal outflow over production values are of the order of unity for the outflows in which 10--60\% of the metals originate from SNe. In the extended dwarf the feedback metals comprise only up to a couple percent. The stellar feedback therefore drags along ambient metals into the outflows at least as much as they inject into the ISM. To conclude, the outflows in dwarf galaxies are driven by the SNe, regulated by the early stellar feedback, and they carry out tens of percents of the produced metals \citep{1999ApJ...513..142M}. On the other hand, only less than 10\% of all the SN and wind material get locked in new stars.

\section{Summary and Discussion}\label{section:summary}

We have analysed the formation of stars and star clusters that have been enriched by stellar feedback in high-resolution hydrodynamical dwarf galaxy simulations. We used two low-mass disk galaxy models at an initial metallicity of $Z=0.00021=0.016$ \zdot, one with a compact gas disk and one with a more extended initial gas distribution. Star formation is modelled using a Jeans-mass dependent threshold without any additional efficiency parameters and the simulated galaxies were shown to agree with the observed relations between gas mass surface density and the SFR surface density in spiral and dwarf galaxies. The simulations sample the full stellar IMF down to 0.08 \mdot{} and ensure mass conservation locally in regions of \mbox{1 pc} in radius. As a result, massive stars form and live most of their lives in bound clusters. The maximum initial masses in individual bound star clusters were also shown to follow a trend between the largest stellar mass and the star cluster mass that differs from the relation obtained from pure stochastic sampling.

We analysed the photometric properties of the star clusters obtained via radiative transfer post-processing of the compact dwarf galaxy after \mbox{500 Myr} of star formation. Clusters were detected using a photometric detection pipeline, motivated by \textit{HST} star cluster surveys. In the best case scenario where the image data is only degraded by stochastic noise and the instrument PSF, the pipeline is able to detect most of the young star clusters in the galaxy at a distance of \mbox{3 Mpc}. On the other hand, the population of low-mass, evolved star clusters is not completely recovered in the synthetic observations and the resulting total bound cluster mass is underestimated by $\sim1/3$. The cluster $M/L_\mathrm{V}$ evolves with the cluster age both due to stellar evolution but also via dynamical evolution of the clusters. Lower-mass stars escape due to the tidal field of the dwarf galaxy, following the mass-loss rates of star clusters in local galaxies. As a result, the $M/L_\mathrm{V}$ values at a given age are lower for lower mass clusters compared to higher mass clusters. The dynamical mass-loss of the clusters may be somewhat underestimated due to the softened gravitational interactions, compared to models that resolve the strong two-body interactions. The limited dynamical range will be addressed in future applications.

The simulations follow the mass of 13 elements released via stellar feedback individually for each feedback channel. Here we followed the combined mass of metals in the SN and wind material and will investigate the evolution of the individual elements in the future. The vast majority of the winds and SNe are produced by massive stars in bound clusters. Overall, the integrated mass of SN ejecta dominate the enrichment, while the number of stars forming out of enriched gas is governed by the continuous stellar wind feedback. We followed the fate of the feedback material being injected into the ISM and concluded that $\sim60\%$ of the SN mass and $\sim35\%$ of the wind mass is either blown out of the galaxies via outflows or reside in the hot gas phase. The metallicity of the outflows is therefore enhanced up to 2--3 times, keeping in mind the initially low pristine gas metallicity. On the other hand, only up to $\sim 6$\% of the SN and $\sim9$\% of the wind mass is locked in new stars. The rest of the material remains in the cold and warm ISM. Stellar winds were therefore shown to be more prone to stay in the ISM of the dwarf galaxies, available for further, enriched star formation. The dwarf galaxy models with and without winds did not show any significant differences in SFR, IMF, ambient feedback densities, outflow rates or mass loading factors. Due to cooling, the momentum imparted by the stellar winds is not fully boosted by the adiabatic expansion of the wind bubble and the energy input of the winds should be considered as a lower limit. However, cooling is expected to be more efficient in a turbulent ISM, therefore the reduced momentum input seen in idealized test scenarios may not in reality affect the results of the galaxy scale simulations.

While the wind and SN feedback, as well as the formation of wind enriched stars, start almost immediately once the first massive stars form, the first SN enriched stars form with a significant delay. In the compact dwarf galaxy, the first SN enriched stars form more than \mbox{$50$ Myr} after the first massive stars, followed by an additional delay of \mbox{10 Myr} in the formation of SN enriched star clusters. In the extended dwarf galaxy the delays are even longer, of the order of \mbox{130 Myr} for stars and almost \mbox{200 Myr} for clusters. We will investigate more intensely star-forming systems in the future to see how these delays behave in environments that are able to form more massive star clusters.

The clusters in our simulations form rapidly over time scales of a few Myr, and the cluster stars can be enriched by various stellar feedback processes. In the context of observed star clusters, the enriched stars could be analogous to the so-called second population (see e.g. \citealt{2018ARA&A..56...83B, 2019A&ARv..27....8G}). Conceptually the process of formation and early enrichment corresponds to the self-enrichment scenario of globular clusters: before SNe occur, new stars can form enriched only with light elements released in stellar winds of massive stars. The majority of the produced SN material is either in the hot ISM or blown out of the galaxy, therefore star formation in gas pockets devoid of SN metals (such as Fe) may also occur in the later phases of galaxy evolution. The tracking of each individual metal from each enrichment channel implemented in our model is therefore crucial for deciphering the origin of each element in the simulated star clusters and will be utilized in our future studies concerning the elemental abundances of globular clusters.

Even though the upper limit for the IMF sampling was set to \mbox{500 \mdot}, the dwarf galaxies in the present study formed only a handful of stars more massive than 100 \mdot. In the compact dwarf galaxy model, only one star formed with high enough initial mass to undergo a PISN. Within our model, more extreme star-forming environments are needed to allow for higher maximum stellar masses and more wind enrichment. Such increase in the number of massive stars will be needed to study the metal enrichment in the context of globular clusters.
The initial surface densities of the simulated clusters reach a few \mbox{$10^3$ \mdot{} pc$^{-2}$} within half-mass radii. For comparison, the surface densities within the effective radius (1--\mbox{10 pc}) of present-day globular clusters reach more than \mbox{$10^3$ \mdot{} pc$^{-2}$} \citep{2011MNRAS.414.3699M} and were orders of magnitude higher at formation \citep{2023MNRAS.520.2180C, 2023ApJ...945...53V}. In future work we will therefore investigate the role of enrichment in more extreme environments such as dwarf galaxy starbursts that are able to form more massive clusters than in the present study. We will test the retention of wind and SN material in globular cluster-forming regions and test the hypothesis of stellar winds of massive stars as the source of multiple populations. In our proposed formation scenario, all the cluster stars form in one generation but include chemically distinct populations of either unenriched, wind enriched and possibly SN enriched stars. Recent simulations of giant molecular cloud collisions indicate that SN ejecta might be able to escape even from the more massive star cluster formation regions \citep{2022ApJ...935...53H}, as long as proper stellar feedback is included. The inclusion of a galactic environment is therefore crucial for assessing if the escaped stellar material can instead enrich other, nearby star-forming regions. Interestingly, recent re-analysis of archival globular cluster data indicates that the so-called first population can, in fact, be enriched in heavy elements \citep{2023A&A...669A..19L} such as SN ejecta. Once the SNe in our simulations couple to the cold ISM (e.g. Fig. \ref{fig:ISM}), even small amounts of the SN material easily overwhelm the pristine metallicity and the wind material. The new observational results are therefore encouraging in light of the results presented here: most of the stars in massive star clusters should be only enriched in light elements (Fig. \ref{fig:ejectaN}), but some enrichment by SNe may still be allowed. 

\section*{Acknowledgements}

We thank Michael Grudi\'c for a constructive referee report, and Patrick Neunteufel, Adrian Hamers, Martyna Chru\'sli\'nska and Pavel Kroupa for useful discussions. 
NL and TN acknowledge the computing time granted by the LRZ (Leibniz-Rechenzentrum) on SuperMUC-NG under
project numbers pn49qi and pn72bu. TN acknowledges support from the Deutsche Forschungsgemeinschaft (DFG, German Research Foundation) under Germany's Excellence Strategy - EXC-2094 - 390783311 from the DFG Cluster of Excellence "ORIGINS". JMH acknowledges the support of the Academy of Finland grant 339127 and the European Research Council via ERC Consolidator Grant KETJU (no. 818930). DS acknowledges the support of the National Science Center (NCN), Poland, under grant number OPUS 2021/41/B/ST9/00757. For the purpose of Open Access, the author has applied a CC-BY public copyright
license to any Author Accepted Manuscript (AAM) version arising from this submission.The computations were carried out at SuperMUC-NG hosted by the Leibniz-Rechenzentrum and the COBRA and FREYA clusters hosted by The Max Planck Computing and Data Facility (MPCDF) in Garching, Germany.

This research made use of \textsc{python} packages \textsc{scipy} \citep{2020SciPy-NMeth}, \textsc{numpy} \citep{2020NumPy-Array}, \textsc{matplotlib} \citep{Hunter:2007}, \textsc{pygad} \citep{2020MNRAS.496..152R} and \textsc{h5py} \citep{collette_python_hdf5_2014}.

\section*{Data Availability}

The data will be made available on reasonable request to the corresponding author.



\bibliographystyle{mnras}




\appendix

\section{Momentum evolution in the wind method}\label{appendix:A}

\begin{figure*}
\includegraphics[width=\textwidth]{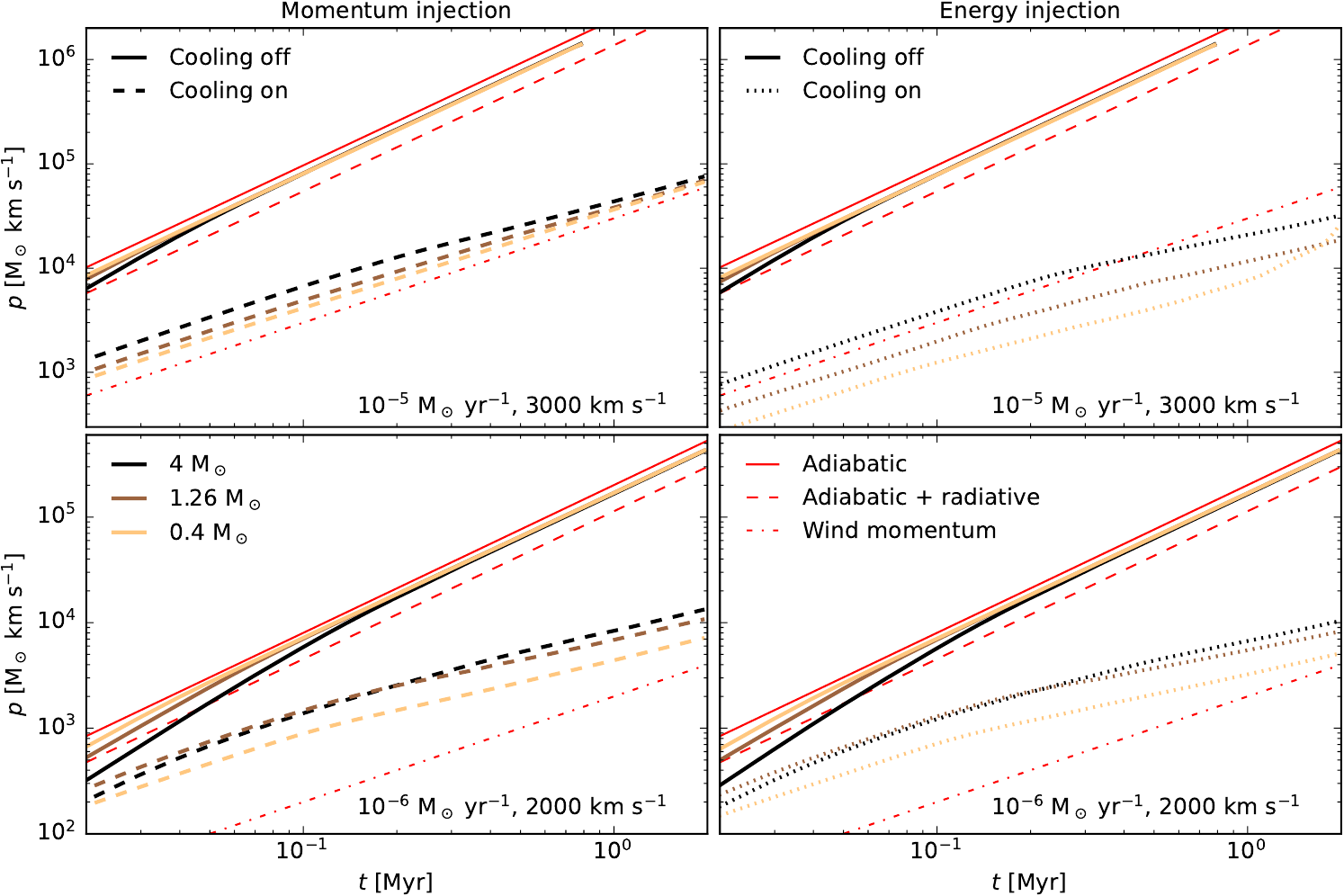}
\caption{The evolution of the momentum in the test runs shown in Fig. \ref{fig:Rshell}. The line colours and styles are the same as in Fig. \ref{fig:Rshell} and the red lines show the analytic solutions corresponding to Eqs. \eqref{eq:energy}--\eqref{eq:momentum}.\label{fig:momentum}}
\end{figure*}

The momentum in the adiabatically expanding wind bubble is dominated by the momentum of the shell
\begin{equation}
    p_\mathrm{shell}=\frac{4\pi}{3} R_\mathrm{shell}^3 \rho_0 \dot{R}_\mathrm{shell} 
\end{equation}
which gives 
\begin{equation}\label{eq:momentum_ad}
    p_\mathrm{shell}=\frac{4}{5\pi} \alpha^4 (L_\mathrm{wind}^4 \rho_0 t^7)^{1/5}
\end{equation}
where $\alpha=0.88$ and $\alpha=0.76$ from the solution of $R_\mathrm{shell}$ in Eqs. \eqref{eq:energy} and \eqref{eq:energy_rad}, respectively. When cooling is efficient, the wind bubble is only driven by the wind momentum $\dot{M}_\mathrm{wind}v_\mathrm{wind}$. We show in Fig. \ref{fig:momentum} the momentum evolution in the hydrodynamical tests shown in Fig. \ref{fig:Rshell} and compare with the three respective analytic solutions.

In Fig. \ref{fig:momentum}, the runs without cooling result in a close match to the evolution given by the momentum in Eq. \eqref{eq:momentum_ad} with $\alpha=0.88$ regardless of injection method. The higher resolution tests without cooling generate momentum faster than the fiducial resolution but all of the runs reach the same solution with time. The runs where cooling is switched on, the wind bubble generates momentum slower. The momentum injection method (left) is able to create a small momentum boost in all test cases. The energy injection method (right), on the other hand, fails to conserve even the input wind momentum when the wind has a high outflow rate and velocity. The momentum injection method is therefore necessary at the resolution of our current dwarf galaxy models.

The evolution in radial momentum is practically identical to the total momentum shown in Fig. \ref{fig:momentum}, therefore the wind is always able to generate and sustain a bubble. If that were not the case, the radial momentum might become negative and lead to unrealistic inflow of matter as demonstrated by \citet{2021MNRAS.508.1768P}. Our model therefore somewhat underestimates the effect of the stellar wind in generating momentum in the uniform ambient medium. However, in turbulent media the efficient mixing should in fact lead to more efficient cooling and bubble evolution that is more like a momentum-driven snowplow \citep{2021ApJ...914...89L}.


\bsp	
\label{lastpage}
\end{document}